\documentclass[a4paper,fleqn]{cas-dc}

\usepackage[numbers]{natbib}
\usepackage{graphicx} 

\usepackage{algorithm}
\usepackage{algorithmic}
\usepackage{hyperref}
\usepackage{xcolor}
\hypersetup{
    colorlinks=true,
    linkcolor=blue,
    filecolor=magenta,      
    urlcolor=cyan,
    citecolor=red
}

\def\tsc#1{\csdef{#1}{\textsc{\lowercase{#1}}\xspace}}
\tsc{WGM}
\tsc{QE}

\begin{document}
\let\WriteBookmarks\relax
\def\floatpagepagefraction{1}
\def\textpagefraction{.001}

\shorttitle{\emph{Robust CNN Multi-Nested-LSTM Framework for Patch-based Multi-Push SWI
}}  


\shortauthors{\emph{M.J.Alam, A.Habib and M.K.Hasan}}  

\title[mode=title]{Robust CNN Multi-Nested-LSTM Framework with Compound Loss for Patch-based Multi-Push Ultrasound Shear Wave Imaging and Segmentation}



%

\author[1]{Md. Jahin Alam\textsuperscript{\dag}}
\author[1]{Ahsan Habib\textsuperscript{\dag}}%
\author[1]{Md. Kamrul Hasan}

\cormark[1]




\affiliation[1]{organization={Department of Electrical and Electronic Engineering, Bangladesh University of Engineering and Technology (BUET)},
            city={Dhaka},
            postcode={1205}, 
            country={Bangladesh}}

\let\printorcid\relax
\cortext[1]{ Corresponding author: khasan@eee.buet.ac.bd (M. K. Hasan).\\
\-\ \-\ \-\ \-\ \-\ \-\ \-\ \-\ \-\ \dag : These authors contributed equally to this work. \\
\-\ \-\ \-\ \-\ \-\ \-\ \-\ \-\ \-\ \emph{Email addresses}: jahinalam.eee.buet@gmail.com (M. J. Alam)}
 

\begin{abstract}
\textit {Objective:} Ultrasound Shear Wave Imaging is a noteworthy tool for $in-vivo$ noninvasive tissue pathology assessment and elasticity estimation for medical applications. State-of-the-art techniques can generate reasonable estimates of tissue elasticity, but high-quality and noise-resiliency in shear wave elastography (SWE) reconstruction have yet to demonstrate advancements. \textit{Approach:} In this work, we propose a two-stage deep-learning pipeline that can not only produce reliable reconstructions from SWE motion data but also denoise said reconstructions to obtain better contrast and lower noise prevailing elasticity mappings. The reconstruction network consists of a Resnet3D Encoder to extract temporal context from the sequential multi-push separated data. The encoded features are sent to multiple feature-level Nested CNN LSTM (Convolutional Long Short-Term Memory) blocks which process them in a temporal attention-guided windowing basis and map the temporal information into the spatial domain using an FFT-based attention module. The spatial features are then decoded into a 2D elasticity modulus map as a form of primary reconstruction. The obtained 2D maps from each multi-push regions are merged and then sent to a dual-decoder denoiser network which denoises the foreground (inclusion) and background features independently before fusing the two. The post-denoiser generates a higher-quality reconstruction as well as an inclusion isolating segmentation mask. Apart from a simple primary reconstruction loss, a multi-objective compound loss is designed to accommodate the denoising, fusing, and mask generation processes. The method is validated on sequential multi-push (both simulation and experimental) SWE motion data with multiple overlapping regions. A patch-based training procedure is also introduced with network modifications to handle data scarcity. \textit {Main Results:} Experiments produce an average 32.66~dB PSNR, 43.19~dB CNR, and 0.996 SSIM in noisy simulation data and an average PSNR of 22.44~dB, CNR of 36.88~dB, and SSIM of 0.943 in a private CIRS049 phantom data, across the test samples. Additionally, IoUs (0.909 and 0.781, respectively) and ASSD (0.227 and 0.863, respectively) were quite satisfactory in the simulated and private data. After comparing with other reported deep-learning approaches, i.e. a Spatio-Temporal CNN and DSWE-Net, our method proves quantitatively and qualitatively superior in dealing with noise influences in simulated and experimental SWE data. \textit {Significance:} From a performance point of view, our deep-learning pipeline has the potential to become utilitarian in the clinical domain.

\end{abstract}

 
\maketitle

\section{Introduction}\label{introduction}

The mechanical properties of soft tissues are highly correlated with tissue pathology. Clinical applications regard these properties, such as the elasticity or stiffness of the tissue, as very significant in assessing the severity of diseases and treatments. Because a proper distinction between different tissue stiffness can lead to identifying cancerous or anomalous regions (liver fibrosis: $>8kPa$, breast tumor: $33.3-80 kPa$) from normal and healthy areas (liver: $<6kPa$, breast fatty tissue: $5-10kPa$, breast parenchyma: $30-50kPa$) \cite{tanter2008quantitative, mueller2010liver, youk2014comparison}. Various techniques have been developed over the past two decades \cite{greenleaf2003selected} with the aim of accurate quantification of the tissue's mechanical properties. Among them, elastography techniques, more specifically strain elastography and shear wave elastography (SWE) techniques have gained recognition in many clinical diagnoses cases for measuring the stiffness property in soft tissues \cite{bamber1999ultrasound, sarvazyan1998shear}. Although it has high contrast, strain-wave imaging provides relative stiffness and is highly operator-dependent \cite{yoon2011interobserver}. Whereas the SWE method can provide absolute stiffness for soft tissues and is reproducible. SWE has been widely used in various clinical non-invasive diagnoses such as breast lesion characterization \cite{Youk}, liver fibrosis \cite{ferraioli2014shear}, and prostate cancer detection \cite{Woo}.


The SWE technique relies on remotely inducing tissue displacement with small-amplitude mechanical perturbations using an acoustic radiation force (ARF) \cite{nightingale2011acoustic, sarvazyan1998shear}. The induced displacements form a transient shear wave (SW) which propagates in the perpendicular direction to the ultrasound ARF beam. Two significant attributes of soft tissues are reasonable to assume \cite{lai2009introduction}: (i) soft tissues are in-compressible, isotropic, linear, and elastic, (ii) the Bulk modulus $\lambda$ of soft tissues are much higher than the shear modulus $(\mu_s)$. Therefore, the following expressions hold true for the stiffness or, Young's modulus (YM), $E$, of such a medium:
\begin{equation}\label{sws}
\mu_s=\rho \cdot v^2
\end{equation}
\begin{equation}\label{sws}
E[Pa] = \mu_s \frac{3 \lambda+2 \mu_s}{\lambda+\mu_s} \cong 3 \mu_s=3 \rho v^2
\end{equation}
Here $\rho$ is the medium density $\left(\approx 1000 \mathrm{~kg} \mathrm{~m}^{-3}\right.$ for soft tissue) and $v$ is the shear wave speed. As evident, accurate measurement of SW motion can provide stiffness estimation of the elastic medium under consideration.

Different techniques have been developed to calculate shear wave speed (SWS) from sequences of tissue displacement imaging. These methods can be divided into either time-domain or frequency-domain approaches. The time domain methods aim to find the SWS by estimating the arrival time between two known locations and are generally known as Time-of-Flight (ToF) methods. The peaks of the displacement (or velocity) signals can be tracked (Time-to-peak, TTP) to find the arrival time \cite{sandrin2003transient,palmeri2008quantifying} or cross-correlating the signal groups can also be employed to do the same \cite{tanter2008quantitative, mclaughlin2006shear, song2012comb, kijanka2019fast}. However, in the presence of artifacts and noise, the wave propagation becomes inconsistent \cite{rouze2012parameters} and ToF methods fall short because of their inability to adapt to uncorrelated noise components. The second approach for SWS estimates the phase velocity of the dominant local wave numbers in the frequency domain. Similar to a 2D cross-correlation, this approach does not rely on the wave direction. The SWS can be calculated using a Fourier-transform-based inversion algorithm \cite{bercoff2004supersonic} with the computation of Laplacian components. This, however, does not assure relevant SWS estimation as the wave components parallel to the ARF beam direction are extremely weak. The utilization of local phase velocity sparsity \cite{kijanka2018local} has been able to produce SW reconstructions from dominant wave numbers (known as Local Phase Velocity Based Imaging, or LPVI). But such an approach requires intensive tuning of the frequency band selection and filter parameters. Also, the method is restricted to the SW motion bandwidth, determined by the medium, acoustic radiation force push beam geometry, and push duration.

The artifacts and noise effects can be handled through the adapting nature of machine learning (ML) and deep learning techniques (DL), as found in the field of compression-based strain wave elastography. Implementations of multi-layer perceptrons (MLP) \cite{zayed2020fast} and encoder-decoder \cite{jush2022deep}-based straightforward mapping as well as Generative Adversarial networks (GAN) \cite{he2020application} and semi-supervised training \cite{tehrani2022bi} for quality improvement have been performed using strain radio-frequency (RF) data structure. However, in SWE, ML-DL techniques have not been explored much. Recent works demonstrate that two studies \cite{ahmed2021dswe, neidhardt2022ultrasound} have investigated SWE deep learning to estimate tissue elasticity parameters. Ahmed et al. \cite{ahmed2021dswe} have shown that deep learning can generate both elasticity as well as segmentation maps from simulated tissue motion data. Additionally, authors have validated that the model trained on simulated data can achieve superior performance on experimental CIRS phantom data compared to the classical SWS estimation method \cite{kijanka2018local} in estimating tissue stiffness mapping. However, they also reported that the noise provided in the simulation data is not sufficient to represent real-world jitter or speckle noise arising during SW acquisition. In contrast, Neidhardt et al. \cite{neidhardt2022ultrasound} have suggested a window-based 3D spatio-temporal Convolutional Neural Network (CNN) to predict local elasticity maps. They have argued that their method can estimate elastic properties on a pixel-wise basis even in the push region, where the classical techniques fail to retrieve any reliable result, especially in noise-inflicted data. But this method needs a relatively large spatial window to produce an acceptable estimation of the SW of a single pixel. Furthermore, training and evaluating this technique to produce a full SWS mapping requires a very long computation time.

\begin{figure*}[t]
  \centering
  \includegraphics[width=0.98\textwidth]{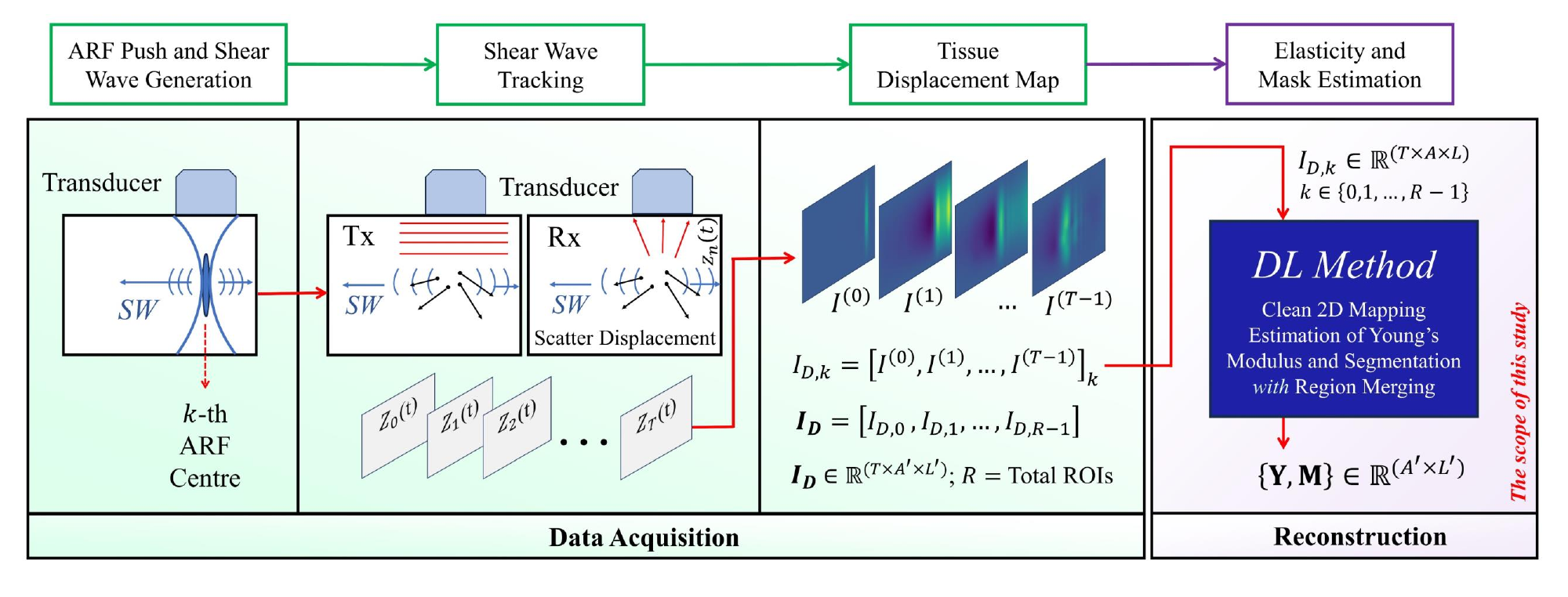}
  \caption{The entire procedural (data acquisition and reconstruction) steps of Shear Wave Elastography.}
  \label{methodology} 
\end{figure*}

Techniques have been developed where, instead of using a single ARF push beam to induce tissue displacement, multiple spatially separated ARF beams are provided. This is performed to cover a larger region of interest (ROI) as well as effectively image the blind areas generated at each ARF center. The notable techniques are comb-push ultrasound shear elastography (CUSE) \cite{song2012comb, song2014fast} and sequential multi-push SWE \cite{inoue2development}. In the CUSE method, the ultrasound transducer is divided into subgroups, transmitting and tracking spatially separated ARF beams simultaneously. The blind area of one ARF is mitigated by the other shear waves; however, the waves constructively and destructively interfere with each other. The sequential multi-push SWE \cite{inoue2development} transmits multiple ARF pushes with tracking times in between. Each push is deployed at a laterally higher distance than the blind area in relation to its ROI, thereby ignoring the blind area. This method is simpler than CUSE due to the transducer not having subgroups. However, the frame rate is lower compared to CUSE due to the multiple tracking sequences. Our method deals with sequential multi-push data where, due to the in-between tracking, each ROI (spatially overlapped) faces one ARF at a time.

In this paper, we propose a two-stage pipeline with a CNN Multi-Nested-LSTM-based reconstruction network and a post-denoiser network for SWE imaging. SWE motion data of each sub-region due to multiple sequentially pushed ARF is passed through the reconstruction network, consisting of an encoder with Resnet3D blocks for spatial and temporal processing. Intermediate encoder features are sent to CNN Nested-LSTM blocks in order to extract the temporal information of the SW-propagation and project it in the feature-level spatial domain. The Nested-LSTM blocks perform windowing along the temporal axis to better isolate the significant time components, and a subsequent temporal-attention module (TAM) selectively weighs and maps the salient features. The magnitude spectrums of the spatially mapped features are then weighted by an FFT-based attention to identify the useful magnitude components. A squeeze-and-excitation \cite{hu2018squeeze} attention-guided decoder reconstructs the 2D stiffness mapping of the motion data. We take into account the data limitation in the medical domain and also propose a patch-based training regime for the reconstruction network to accommodate for better training and reconstruction of each region. The resulting 2D reconstructions of the regions are spatially merged to obtain the reconstruction of the entire ROI. In the next stage, a post-denoiser network is introduced to clean the total ROI map by processing the foreground (inclusion) and background separately. This is achieved with a 2D-encoder pipeline shared by two identical 2D decoders. A `Fusion' module is used to register the foreground and background into a denoised modulus estimate from the feature level. The module also generates a segmentation mask for the inclusion region. The primary reconstruction task is performed with a simple mean-absolute-error. The denoiser task is supervised to be noise resilient by employing a multi-objective compound loss consisting of foreground and background regional losses, a fusion loss, a total-variational loss, and an IoU loss. The method is trained and evaluated on sequential multi-push-generated data (both simulation-based and CIRS phantom-based), with motion from each region representing a single data sample. The results analyzed from the datasets demonstrate the superiority of our deep learning pipeline in contrast to the previously reported DL works.

\section{Materials and Method} \label{methods}

\subsection{Problem Formulation} \label{problem_form}

An ARF induces displacement fields in a soft tissue medium in the form of shear waves that travel perpendicularly to the push beam. The propagating SW (i.e., tissue displacement) is captured with the help of plane wave imaging \cite{couture2012ultrasound}. In our study, we consider the motion of scatterers along the axial dimension in the axial and lateral planes, which corresponds to the direction of the tracking beam propagation. When a single-point scatterer is located along the tracking line, the captured RF signal by an ultrasound transducer can be represented by the following model:

\begin{equation} \label{x_t}
\begin{aligned}
z_0(t) &= h\left(t - \tau_0\right) \\
&= A\left(t - \tau_0\right) \cos\left(\omega_c\left(t - \tau_0\right)\right), \;\;  {z}_0(t) \in \mathbf{Z}_{\mathrm{0}}(t)
\end{aligned}
\end{equation}
Here, $\omega_c$ denotes the angular frequency of the ultrasound carrier, $\tau_0$ represents the travel time between the transducer and the point scatterer, $h$ is the impulse response, and $A$ is the real envelope of the received pulse. $\mathbf{Z}_{\mathrm{0}}(t)$ represents the RF signal of all the points in the region of interest. When a subsequent measurement is conducted after a slight displacement of the scatterer, the recorded signals can be expressed as
\begin{equation}
\begin{aligned}
 z_n(t) &= S\left(t - \tau_n\right) = A\left(t - \tau_n\right) \cos\left(\omega_0\left(t - \tau_n\right)\right)
\end{aligned}
\end{equation}
where,
\begin{equation}
\begin{aligned}
 z_n(t) &\in \mathbf{Z}_n(t),\;and, n = 0, 1, \ldots, T-1
\end{aligned}
\end{equation}
Here, $\tau_n$  signifies the propagation time between the transducer and the axially displaced point scatterer with respect to its initial position at $\tau_0$. The displacement of the scatterer can be directly related to the difference in time delays using the equation
\begin{equation}\label{del_r}
\begin{aligned}
\Delta r_n &= c\left(\tau_{n} - \tau_{n-1}\right) \\
&= c\Delta\tau_n, \;\; n = 0, 1, \ldots, T-1 
\end{aligned}
\end{equation}
where $c$ represents the velocity of sound, the value of $\Delta\tau_n$ can be determined by locating the peak of the cross-correlation between $z_{n-1}(t)$ and $z_n(t)$. Then the absolute displacement with respect to the initial position of single-point scatter at time $T$ can be calculated as
\begin{equation}\label{del_r}
r_{n} = \sum_{i=0}^{n} \Delta r_i 
\end{equation}
This concept of tracking displacements for a single-point scatterer can be extended to a large number of uniformly distributed scatterers within a 2D ROI. If there are a total of $A$ and $L$ spatial points along the axial and lateral axis, respectively, in the regions, we can denote the absolute displacement of a single point scatterer located at (axial=$a$, lateral=$l$) coordinates to be $r_{n}^{(a,l)}$. Therefore, the entire 2D region can be expressed as

\begin{equation}
I^{(n)}= \begin{bmatrix}
r_{n}^{(0,0)} & r_{n}^{(0,1)} & \cdots & r_{n}^{(0,L-1)} \\
r_{n}^{(1,0)} & r_{n}^{(1,1)} & \cdots & r_{n}^{(0,L-1)} \\
\vdots & \vdots & \ddots & \vdots \\
r_{n}^{(A-1,0)} & r_{n}^{(A-1,1)} & \cdots & r_{n}^{(A-1,L-1)}
\end{bmatrix}
\end{equation}
where $I^{(n)}$ represents the 2D tissue displacement data with spatial dimensions of $(A \times L)$ at the $n$-th time frame. As we are dealing with sequential multi-push SWE, there will be multiple spatially overlapping regions. For the $k$-th region, by tracking the displacement data for all the frames in a single imaging sequence, we can get 3D volume data, which is generated by stacking 2D particle velocity maps for the entire ROI at different time steps:
\begin{figure*}[t]
  \centering
  \includegraphics[width=1\textwidth]{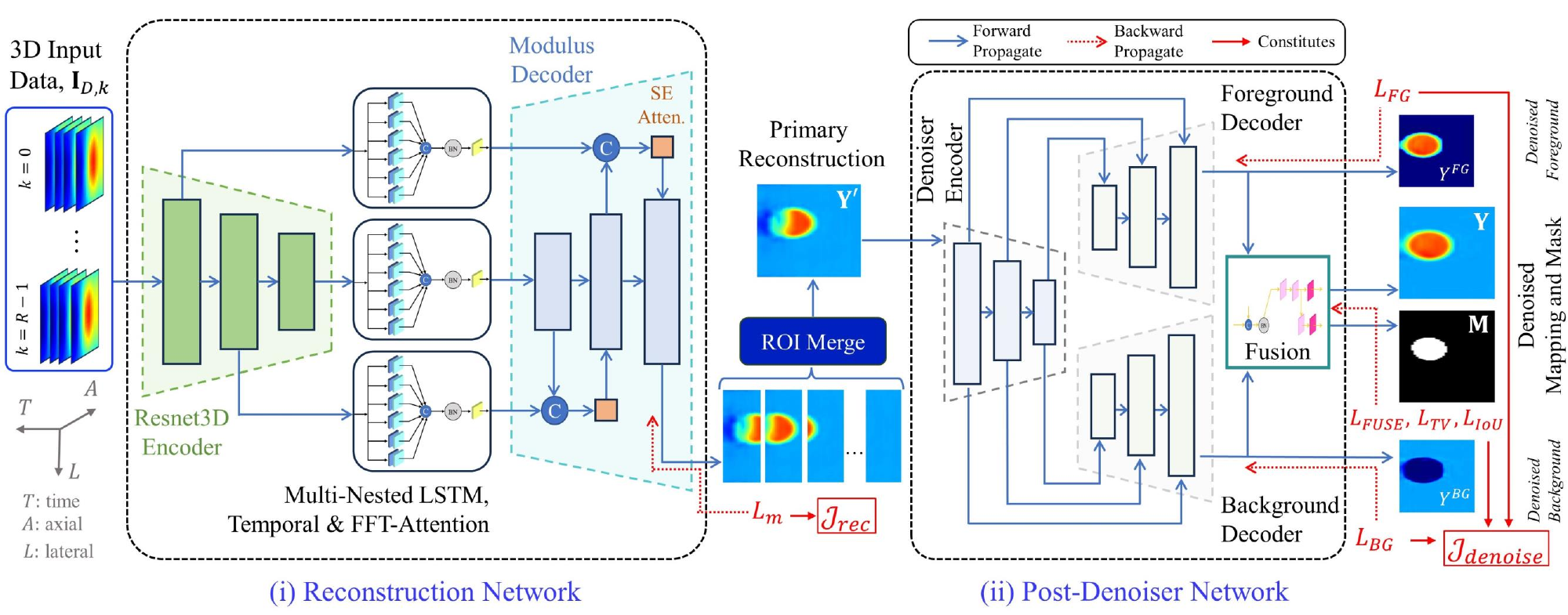}
  \caption{Graphical Abstract of the complete network architecture and the associated Compound Loss functions.}
  \label{Complete_Network_block_diagram} 
\end{figure*}

\begin{equation} \label{Region_in_vector}
\begin{split} 
\textbf{I}_{D,k} = \left[I_D^{(0)}, I_D^{(1)}, \dots, I_D^{(T-1)}\right]_k \;, & \;\; \textbf{I}_{D,k} \in \mathbb{R}^{T \times A \times L } \\
& \;\; k \in \{0,1,...,R-1\}
\end{split}
\end{equation}
Here, $R$ denotes the total regions present in the data under an entire ROI. The task of SWE image reconstruction and lesion segmentation can be reduced to the transformation from displacement data, $\textbf{I}_{D,k}$, to a 2D Young's modulus map, $\mathbf{Y}$, and a binary mask, $\mathbf{M}$. The corresponding output of each ROI is separately produced, as the ARF pushes are separate. Being overlapping regions, the 2D reconstruction and masks can be merged with spatial windowing. The merged primary reconstruction may be suboptimal due to the presence of noise in the data. Therefore, another transformation can be introduced to clean it as well as produce the segmentation mask. The scope of our study in schematic form is shown in figure \ref{methodology} alongside the data previously described data acquisition process. 

In our approach, the aforementioned transformations are learned in two stages. The first transfer function, $\mathcal{F}_{Recon}$ (i.e., primary reconstruction network) learns to map an intermediate Young's elasticity modulus directly from displacement data. The second transfer function $\mathcal{F}_{Denoise}$ further refines the intermediate reconstruction to produce a more robust estimation as well as outputs a corresponding segmentation mask (see detailed architecture in figure \ref{Complete_Network_block_diagram}). The process can be defined mathematically as follows: 
\begin{equation}\label{YM1}
\textbf{{Y}}'_k= \mathcal{F}_{Recon}\left(\textbf{I}_{D,k};\; \Theta_\mathcal{P}\right), \;\; \textbf{Y}_k' \in \mathbb{R}^{A \times L } 
\end{equation}
\begin{equation}\label{YM2}
\textbf{Y}'= \mathcal{S}_m \left(\textbf{{Y}}'_0, \textbf{{Y}}'_1, ..., \textbf{{Y}}'_{R-1} \right), \;\; \textbf{Y}' \in \mathbb{R}^{A' \times L' }
\end{equation}
\begin{equation}\label{YM3}
\left\{\textbf{Y}, \textbf{M} \right\}   = \mathcal{F}_{Denoise}\left(\textbf{Y}';\; \Theta_\mathcal{D} \right), \;\; \{\textbf{Y}, \textbf{M}\} \in \mathbb{R}^{A' \times L' } 
\end{equation}
Here, $\Theta_\mathcal{P}$ represents the learnable parameters of the primary reconstruction network and $\Theta_\mathcal{D}$ represents the learnable parameters of the denoiser network. After merging the reconstructions from each ROI, denoted by $\mathcal{S}_m(\cdot)$, $\textbf{Y}'$ is obtained as the complete primary reconstruction. And, $\textbf{Y}$, $\textbf{M}$ are the cleaned version of the elasticity mapping and the segmented area, respectively.

\subsection{Proposed Network Architecture} \label{Proposed Network Architecture}

The network architecture of our approach consists of two stages: (i) Reconstruction Network which takes a propagating SW displacement profile as input and outputs an intermediate reconstruction, and (ii) Post Denoiser Network to further refine the reconstructed output of the previous block. The graphical summary of our proposed network is shown in figure \ref{Complete_Network_block_diagram}.

\subsubsection{Reconstruction Network}
The reconstruction network was designed to estimate Young's Modulus (YM) profile and lesion mask from a given volumetric SW displacement data. In order to achieve this we incorporated a sophisticated network architecture consisting of several key modules. 

\begin{figure*}[t]
  \centering
  \includegraphics[width=1\textwidth]{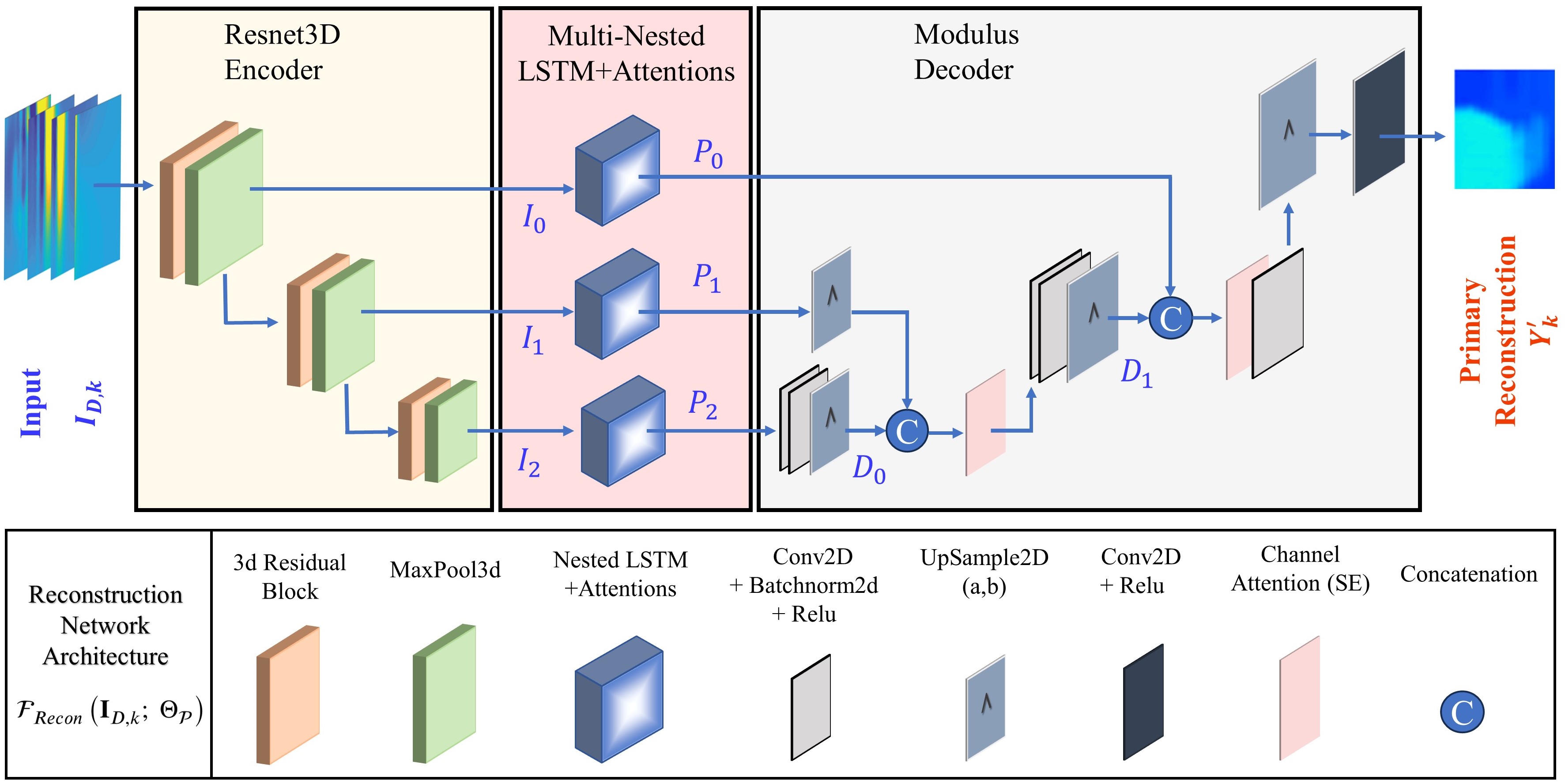}
  \caption{Detailed architecture of the proposed Encoder-Decoder based reconstruction network. }
  \label{Network_block_diagram} 
\end{figure*}

The first important element is a 3D residual spatio-temporal encoder block which serves to encode the 3D volumetric SWS data into a more robust feature space. In the next step, the encoded feature space is passed to a novel multi-nested convolution-LSTM layer to convert the 3D temporal features space to 2D spatial features, matching the dimension number of the required YM mapping. Finally, these spatial features are passed to a 2D decoder block to get the stiffness estimate. The detailed block diagram of our reconstruction network is shown in figure \ref{Network_block_diagram}.

\textbf{Residual Spatio-Temporal Enocoder}: 

We propose a 3D convolution encoder to encode propagating SW information from tissue displacement changes in multiple frames. The convolution encoder contains $3$ stages of residual blocks. These blocks are based on $3$D Resnets \cite{hara2017learning} to benefit from the inherent skip connections \cite{hochreiter1998vanishing}. Each convolution layer in our network uses $(3 \times 3 \times 3)$ kernel size followed by a batch normalization layer. The input displacement map, $\mathbf{I}_{{D,k}} \in \mathbb{R}^{B \times 1 \times T \times A \times L}$, is passed through three residual blocks where the number of kernels used in the three blocks is $16$, $32$ and $64$, respectively, from the first block to the last block (index: $B$=batch, $C$=channel, $T$=temporal, $A$=axial, $L$=lateral). After the first and second encoding stages, we use  $(2 \times 2 \times2)$ max pooling operation to downsample both spatial and temporal information. The output of the first stage be denoted as $\mathbf{I}_{\mathrm{0}} \in \mathbb{R}^{B \times C \times \frac{T}{2} \times \frac{A}{2} \times \frac{L}{2}}$ and the second stage is denoted as $\mathbf{I}_{\mathrm{1}} \in \mathbb{R}^{B \times \mathrm{2C} \times \frac{T}{4} \times \frac{A}{4} \times \frac{L}{4}}$. After the final encoding block, we perform $(1 \times 2 \times 2)$ maxpooling operation along the spatial dimension leaving the temporal resolution intact which makes the final encoded feature map $\mathbf{I}_{\mathrm{2}} \in \mathbb{R}^{B \times \mathrm{4C}  \times \frac{T}{4} \times \frac{A}{8} \times \frac{L}{8}}$, where $C=16$. This serves two purposes. Firstly, the initial two stages reduce both temporal and spatial dimensions. This provides the later CNN layers with high fidelity spatio-temporal features that assist in encoding better feature representation. Secondly, by using $(1 \times 2 \times 2)$ max pooling after the final encoder block, sufficient distinguishable temporal information is preserved. This assists the final LSTM block in producing more refined correlated feature space between the consecutive time steps, which is crucial for stiffness estimation. We can represent the encoder networks as follows:

\begin{equation}
\mathbf{I}_{\mathrm{0}} = max_{2\times2\times2}\left(\mathcal{F}_0 \left(\mathbf{I}_{{D,k}}, {W_{r0}}\right) + \mathbf{I}_{{D,k}}\right), \;\;  \mathbf{I}_{\mathrm{0}} \in \mathbb{R}^{C \times \frac{T}{2} \times \frac{A}{2} \times \frac{L}{2}}
\end{equation}

\begin{equation}
\mathbf{I}_{\mathrm{1}} = max_{2 \times 2 \times 2}\left(\mathcal{F}_1\left(\mathbf{I}_{\mathrm{0}}, {W_{r1}}\right) + \mathbf{I}_{\mathrm{0}}\right), \;\; \mathbf{I}_{\mathrm{1}} \in \mathbb{R}^{\mathrm{2C} \times \frac{T}{4} \times \frac{A}{4} \times \frac{L}{4}}
\end{equation}
\begin{equation}
\mathbf{I}_{\mathrm{2}} = {max_{1\times2\times2}}\left(\mathcal{F}_2\left(\mathbf{I}_{\mathrm{1}}, {W_{r2}}\right) + \mathbf{I}_{\mathrm{1}}\right), \;\; \mathbf{I}_{\mathrm{2}} \in \mathbb{R}^{\mathrm{4C} \times \frac{T}{4} \times \frac{A}{8} \times \frac{L}{8}}
\end{equation}
where $\mathcal{F}_0$, $\mathcal{F}_1$, and $\mathcal{F}_2$, represents the three residual blocks, and  ${W_{r0}}$, ${W_{r1}}$, and ${W_{r2}}$ are the weights that need to be optimized for each block.

\vspace{2\baselineskip}

\textbf{Multi-Nested LSTM Temporal Processing}:

\begin{figure*}[h]
\centering
\includegraphics[width=0.98\textwidth]{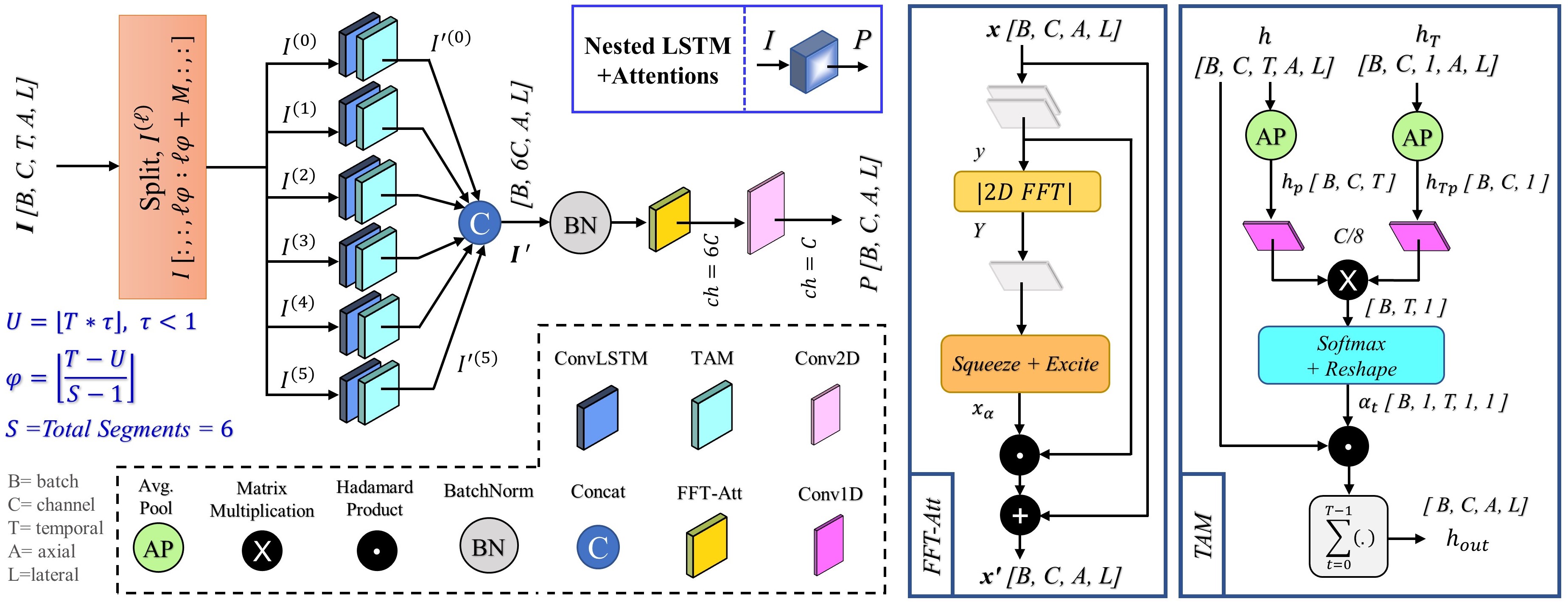}
\caption{Nested-LSTM structure with Temporal and FFT-based Attention.}
\label{nested_lstm} 
\end{figure*}

In our network, we used nested-LSTM temporal blocks after each maxpooling stage. The purpose of incorporating LSTM in such a way as to learn and correlate temporal features from multi-resolution spatial feature space. There are a total of three nested-LSTM blocks (therefore named, Multi-Nested LSTM) after the three max-pooling stages. Each nested-LSTM block, shown in figure \ref{nested_lstm}, is comprised of six parallel paths ($S=6$) to perform windowing of the features $\textbf{I}_0, \textbf{I}_1,$ and $ \textbf{I}_2$ across the temporal axis. Each input feature is split across the time axis, having $\tau$-fraction of the original size ($T$) as well as possessing overlap ($\varphi$) with each other. We split the features into feature segments ($\mathscr{l}=0,1,...,S-1$) to extract necessary information from each motion segment. Shear wave propagation may start early or be delayed, making it unsuitable to process all temporal-frames equally. Learnable ConvLSTMs are used to extract necessary information from each segment, and parallel processing with temporal windows lowers variance among estimations (inspired by the Welch method \cite{so1999comparison} developed for spectral analysis).

The ConvLSTMs have 3 convolution LSTM layers \cite{shi2015convolutional} stacked together sequentially. Following each ConvLSTM, the proposed Temporal Attention Module (TAM) is utilized. The six outputs are concatenated and batch-normalized (BN) and faced with a Fourier Transformation (FFT)-based Attention. The process can be expressed as

\begin{equation} \label{ConvLSTM_TAM}
      \mathbf{I}_i^{'(\mathscr{l})} = \mathcal{A}_i^{tam(\mathscr{l})}\left(\mathbf{ConvLSTM}\left(\mathbf{I}_i^{\mathscr{l}}, {W}_{ci}^{(\mathscr{l})}\right), {W}_{ti}^{(\mathscr{l})}\right)
\end{equation}
\begin{equation}
      \mathbf{I}_i^{'} = concat \left( \left\{\mathbf{I}_i^{'(\mathscr{l})}: \mathscr{l} \in \left[0,S-1\right] \right\}\right)
\end{equation}
\begin{equation}
      \mathbf{P}_i = \mathcal{F}_{conv}\left(\mathcal{A}_i^{fft}\left(BN\left(\mathbf{I}_i^{'}\right)\right), {W}_{nested-i}\right), \;  i = 0,1,2
\end{equation}
The $\mathcal{A}^{tam}$ and $\mathcal{A}^{fft}$ denote the implementation of TAM and FFT-based attention, respectively. After introducing the input features  $\textbf{I}_0, \textbf{I}_1,$ and, $ \textbf{I}_2$, we obtain the time reduced features $\textbf{P}_0 \in \mathbb{R}^{B \times \mathrm{C} \times \frac{A}{2} \times \frac{L}{2}}, \textbf{P}_1 \in \mathbb{R}^{B \times \mathrm{2C} \times \frac{A}{4} \times \frac{L}{4}},$ and $ \textbf{P}_2 \in \mathbb{R}^{B \times \mathrm{4C} \times \frac{A}{8} \times \frac{L}{8}}$, respectively.




\vspace{3mm}
\textbf{Temporal Attention Module (TAM)}: 

In order to provide further selectivity of the temporal components found from the ConvLSTM layers in equation (\ref{ConvLSTM_TAM}), we introduce a temporal attention module (TAM: $\mathcal{A}^{tam}$) shown in figure \ref{nested_lstm}. The temporal attention mechanism enables the model to dynamically focus on the most important features within all the hidden state outputs from the sequences and thus improve the temporal context. For this, a weight vector $\alpha_t \in \mathbb{R}^{(B \times 1 \times T \times 1 \times 1)}$  is generated to be applied across the temporal dimension. Assuming an input feature vector  $h \in \mathbb{R}^{(B \times C \times T \times A\times L)}$ for TAM, the weight is calculated as  
\begin{equation}
\alpha_t = \text{{softmax}} \left[\mathcal{F}_{conv}\left(h_{T},W_T\right)^\mathcal{T} \otimes \mathcal{F}_{conv}\left(h_{Tp}, W_p\right)\right]^{reshape}
\end{equation}
where,
\begin{equation}
h_p = avg_{2\times2}(h) \in \mathbb{R}^{(B \times T \times C )}, \; h_{Tp}= h_p[:,T-1,:]
\end{equation}
Here, $avg_{2\times2}$ and $\otimes$ indicate 2D average pooling and matrix multiplication, respectively. The subscript $t$ and $\mathcal{T}$ indicate the time-index of $\alpha_t$ weighting parameters and transposition operation, respectively. Finally, the output is taken as
\begin{equation}
h_{out} = \sum_{t=0}^{T-1} h \cdot \alpha_t, \quad h_{out} \in \mathbb{R}^{(B \times C \times A \times L)}
\end{equation}

\vspace{3mm}
\textbf{Fourier Transform Based Attention (FFT-Att)}: 

The proposed FFT-based attention, depicted in figure \ref{nested_lstm}, takes advantage of the original Squeeze-and-Excite (SE) operation, but converts the spatial domain (axial, $A$ and lateral, $L$) into the corresponding 2D frequency domain beforehand. The motivation for using the Fourier transform is to tap into the magnitude spectrum where all the sparsity and intensity information of the feature space reside. The six groups of features $\mathbf{I}_i^{'(\mathscr{l})}$ contain different information as they are generated from different time spans. Among them, the feature having more information should have sparser values and dominant magnitude spectrum components. These components captured with a global average pooling layer (GP) that produces intensity vectors.

To produce the attention-map, an input feature, $y\in \mathbb{R}^{(B \times C \times A \times L)}$, is pre-conditioned twice and then its magnitude spectrum is calculated. The channel-wise SE operation is performed to obtain a weight, $x_{\alpha}\in \mathbb{R}^{(B \times C \times 1 \times 1)}$ which is used to apply attention on the input. The steps are as follows:

\begin{equation}
y = \mathcal{F}_{conv, BN, Relu}^{\; \times 2}(x, {W_{12}})
\end{equation}

\begin{equation}
\mathscr{Y}[B,C,u,v] = \left| \sum_{a=0}^{A-1} \sum_{l=0}^{L-1} y[B,C,a,l]e^{-j(\frac{ua}{A}+\frac{vl}{L})} \right|
\end{equation}

\begin{equation}
x_{\alpha} = \mathcal{F}_{SE}\left(\mathcal{F}_{conv, BN, Relu}\left(\mathscr{Y}, {W}_{3}\right), {W}_{SE}\right)
\end{equation}

\begin{equation}
x' = x+ x\;.\;x_{\alpha} 
\end{equation}
Here, $\mathcal{F}_{SE}$ is the channel-wise squeeze-and-excite operation.

\vspace{3mm}
\textbf{Attention-Guided Decoder Block}: 

The decoder module is designed to produce the primary elasticity estimation. We named it "modulus decoder" keeping with the analogy of the task. There are three stages of convolution block along with upsampling to reconstruct the modulus estimation. In each stage, a $(2 \times 2)$ upsampling is followed after convolution operations. This ensures the output to result in the original spatial shape. 

The outputs from the Nested-LSTM blocks ($\mathbf{P}_{2-i} \in \mathbb{R}^{B \times 2^{2-i}C \times \frac{A}{2^{2-i}} \times \frac{L}{2^{2-i}}}, i=0,1,2$) are concatenated with the previous decoder level feature, $\mathbf{D}_{i-1}$. Then it is passed to a Conv2D+BN+ReLu layer ($\mathcal{F}_{di}$) and spatially upsampled. The upsampled versions are concatenated with the upper-level convLSTM block ($\mathbf{P}_{1-i} \in \mathbb{R}^{B \times 2^{1-i}C \times \frac{A}{2^{1-i}} \times \frac{L}{2^{1-i}}}$) and passed through SE-attention blocks to give relevant weight to their most important channels. The entire process can be described as

\begin{equation}
    \mathbf{D}_i = up_{2 \times 2}\left( \mathcal{F}_{di}\left(\mathbf{D}_{i-1}, {W}^{d}_{i}\right)\right), \;\; i=0,1,2
\end{equation}
\begin{equation}
  \mathbf{D'}_i = concat \left(\mathbf{D}_{i}, \mathbf{P}_{1-i}\right), \;\;  i=0,1
\end{equation}
\begin{equation}
  \mathbf{D}^{SEatt}_i = \mathcal{F}^{i}_{SEatt}\left(\mathbf{D'}_i , W^{i}_{SEatt}\right), \;\;  i=0,1
\end{equation}
Here, the decoder architecture begins with $\mathbf{D}_{-1}=\mathbf{P}_2$. Spatial upsampling is performed 3 times ($i=0,1,2$) as opposed to the other operations which take place twice ($i=0,1$). As a result, $\mathbf{D}_2 \in \mathbb{R}^{C \times 1 \times A \times L}$ is finally obtained. It is passed to a Conv2D layer followed by a RELU activation to obtain the primary modulus reconstruction $\textbf{Y}_k'$ as
\begin{equation}
    {\textbf{Y}_k'} =\mathcal{F}_{conv,Relu}\left(\mathbf{D}_2, {W}_{conv}\right),\;\; \mathbf{Y}_k' \in \mathbb{R}^{B \times 1 \times A \times L}
\end{equation}

With a large SWE data pool, training the primary reconstruction network to generate $\textbf{Y}_k'$ is possible. However, data samples are limited in the medical domain. As a result, an alternative implementation of the reconstruction network is proposed in the following subsection.

\subsubsection{Patch-based Training} \label{Patch_based_training_section}
A DL network requires sufficient data samples to learn the accurate mapping between 3D motion frames to 2D stiffness reconstruction, which is otherwise not available in the medical domain, especially $in-vivo$ SWE data. A probable solution might be to use each spatial pixel in the motion structure with the size $\mathbb{R}^{B \times 1 \times T \times 1 \times 1}$ and perform pixel-by-pixel reconstruction, similar to the work of Neidhardt er al. \cite{neidhardt2022ultrasound}. But, pixel-by-pixel reconstruction can be highly susceptible to noise. Therefore, we aim to reconstruct spatial patches using motion patches. This is because every pixel is not affected by noise to the same extent and patches instead of single pixels will allow the model to lower the overall noise variance.

With this in view, the architecture of the proposed reconstruction network is modified so that the data limitation is compensated. When faced with scarce samples, instead of utilizing $\mathbf{I}_{{D,k}} \in \mathbb{R}^{B \times 1 \times T \times A \times L}$, we take a spatial patch $\mathbf{I}_{{Dp,k}} \in \mathbb{R}^{B \times 1 \times T \times A_p \times L_p}$ where $A_p<A, L_p<L$. Since the spatial dimension of the input is reduced, the perceptive field of the network will decrease as well, and mapping $\mathbf{I}_{{Dp,k}} \in \mathbb{R}^{B \times 1 \times T \times A_p \times L_p}$ into the corresponding $\textbf{Y}'_p \in \mathbb{R}^{B \times 1 \times A_p \times L_p}$ will become difficult. Therefore, the network is trained to produce a smaller 2D reconstruction from $\mathbf{I}_{{Dp,k}}$, specifically $\textbf{Y}'_p \in \mathbb{R}^{B \times 1 \times \left\lceil \frac{A_p}{3} \right\rceil \times \left\lceil \frac{L_p}{2} \right\rceil -1}$. The architecture of the encoder, multi-nested LSTM, and decoder are to be modified accordingly. The changes are implemented using convolutional padding and fractional upsampling. The feature sizes for such a patch-based training are shown in table \ref{Patch_based_train}. If, for instance, the reconstructions within the FOV are taken with no spatial overlapping, the training data will increase by 6-fold within the patch alone; because $\textbf{Y}'_p$ covers almost one-third axial and half the lateral dimension of the input patch. This implies that data can be increased significantly for training the reconstruction network. As a result, we can expect to obtain generalized and precise modulus mappings in the presence of noise. 

\begin{table}[h]
    \centering
    \caption{Feature sizes during patch-based training}
    \label{Patch_based_train}
    \begin{tabular}{cc}
    \hline
        \textbf{Feature} & \textbf{Size} \\  \hline
         $\mathbf{I}_{Dp,k}$ & ${B \times 1 \times T \times A_p \times L_p}$ \\ [1mm]
         $\mathbf{I_{\mathrm{0}}}$ & ${B \times C \times T/2 \times \left\lceil A_p/3 \right\rceil \times \left\lceil L_p/2 \right\rceil}$ \\ [1mm]
         $\mathbf{I_{\mathrm{1}}}$ & ${B \times 2C \times T/4 \times \left\lceil A_p/9 \right\rceil \times \left\lceil L_p/4 \right\rceil}$ \\ [1mm]
        $\mathbf{I_{\mathrm{2}}}$ & ${B \times 4C \times T/4 \times \left\lceil A_p/9 \right\rceil \times \left\lceil L_p/8 \right\rceil}$ \\ [1mm]
        $\mathbf{P_{\mathrm{0}}}$ & ${B \times C \times \left\lceil A_p/3 \right\rceil \times \left\lceil L_p/2 \right\rceil}$ \\ [1mm]
        $\mathbf{P'_{\mathrm{0}}}$, $\mathbf{D_{\mathrm{0}}}$ & ${B \times C \times \left\lceil A_p/3 \right\rceil \times \left\lceil L_p/2 \right\rceil -1}$ \\ [1mm]
        $\mathbf{P_{\mathrm{1}}}$, $\mathbf{D_{\mathrm{1}}}$ & ${B \times 2C \times \left\lceil A_p/9 \right\rceil \times \left\lceil L_p/4 \right\rceil}$ \\ [1mm]
        $\mathbf{P_{\mathrm{2}}}$, $\mathbf{D_{\mathrm{2}}}$ & ${B \times 4C \times \left\lceil A_p/9 \right\rceil \times \left\lceil L_p/8 \right\rceil}$ \\ [1mm]
        $\mathbf{Y}'_p$ & ${B \times 1 \times \left\lceil A_p/3 \right\rceil \times \left\lceil L_p/2 \right\rceil -1}$ 
         \\ [1mm] \hline
        
    \end{tabular}
    
\end{table}

\begin{figure*}[t]
\centering
\includegraphics[width=.98\textwidth]{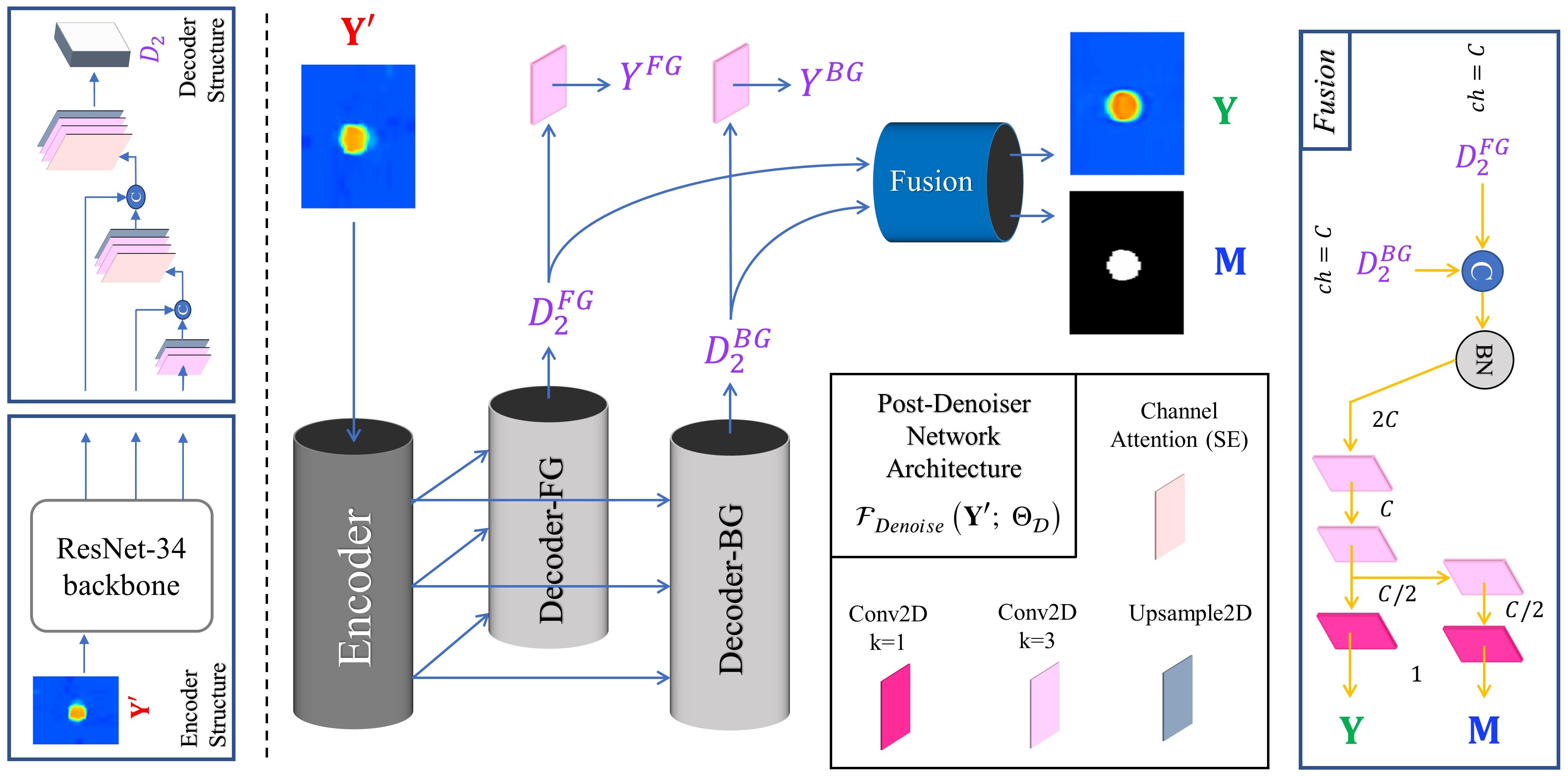}
\caption{Pipeline of the dual purpose post-denoiser network with the fusion block.}
\label{post_denosing_block} 
\end{figure*}

All the 2D overlapping prediction patches, $\textbf{Y}'^{(a,l)}_p$, are first obtained within the $k$-th region ($a,l$ are the top-left axial and lateral coordinates of the patches, respectively). The reconstruction of that region can be produced using 2D windows. The process can be described as

\begin{equation}\label{patch_system_0}
\textbf{{Y}}'^{(a,l)}_p= \mathcal{F}_{Recon}^{patch}\left(\textbf{I}_{Dp,k}^{(a,l)};\; \Theta_\mathcal{P}^{patch}\right), \;\; \textbf{{Y}}'_p \in \mathbb{R}^{A \times L } 
\end{equation}
\begin{equation} \label{patch_system_1}
\mathbb{P}_k = \begin{bmatrix}
\textbf{Y}'^{\left(a_1,l_1\right)}_p  & \cdots & \textbf{Y}'^{\left(a_1,L-1-\frac{L_p}{2}\right)}_p \\
\textbf{Y}'^{\left(a_2,l_1\right)}_p  & \cdots & \textbf{Y}'^{\left(a_2,L-1-\frac{L_p}{2}\right)}_p \\
\vdots  & \ddots & \vdots \\
\textbf{Y}'^{\left(A-1-\frac{A_p}{3},l_1\right)}_p  & \cdots & \textbf{Y}'^{\left(A-1-\frac{A_p}{3},L-1-\frac{L_p}{2}\right)}
\end{bmatrix}_{\;k} 
\end{equation} 
\begin{equation} \label{patch_system_2}
    \textbf{Y}'_k= \mathcal{W}\left(\mathbb{P}_k\right), \;\; k \in {0,1, ..., R-1}
\end{equation}
Here, $\mathcal{W}(\cdot)$ indicates the implementation of 2D windowing (i.e., Tukey windows) to appropriately perform overlapped sum of the patches and produce the entire ROI. In this manner, after procuring the primary reconstruction $\textbf{Y}'_k$,  equations (\ref{YM2}) and (\ref{YM3}) can be used to finally obtain the denoised output and mask.

\subsubsection{Post-Denoiser Network}
A good reconstruction of elasticity modulus from the propagating shear wave depends on the coherent shear wave tracking. But factors such as interference, thermal noise, speckle noise, clutter noise, reflection, and motion artifacts, etc. can adversely affect the SWS estimation, hampering the reliability of stiffness estimations \cite{deffieux2011effects, elegbe2013single, nitta2021review}. Various denoising schemes have been experimented with to alleviate the effect of noise in different studies \cite{cui2019pet, jifara2019medical, el2022efficient}. However, it is very difficult to prevent the noise effects from entering into data with such pre-processing techniques. Deep learning methods, especially CNNs, have showcased reliability in extracting information from complex features and can inherently handle denoising. The primary reconstruction may still be affected by sudden noisy inputs and residual reconstruction noise. To mitigate these effects and enhance the robustness of YM estimation in this work, we introduce a post-denoiser network.

The unique attribute of our post-denoiser is that it handles the inclusion foreground and non-inclusion background separately. The modulus of an inclusion region and a background area are nominally different. Furthermore, the stiffness of the probable inclusion region is more significant from a clinical point of view. Therefore, our post-denoiser has been designed with dual purposes: cleaning both regions separately and then registering the two regions to generate a complete as well as clean 2D modulus mapping. The proposed architecture of it is depicted in figure \ref{post_denosing_block}. 

The post-denoiser network contains an encoder, but dual decoders. The dual decoder pipelines are designated for the inclusion (foreground) and background. We used a ResNet-34 backbone as our encoder pipeline. Each decoder architecture is designed with a structure similar to that employed in the reconstruction network. Subsequently, we introduce a `Fusion' block which takes the foreground and background features from the decoders as inputs and generates a complete clean ROI mapping. The single encoder pipeline is proposed to be sufficient in the denoising both foreground and background, as the $\textbf{Y}'$ is a 2D image which has lower complexity than the 3D input $\textbf{I}_{D,k}$.

\textbf{Encoder Pipeline:} 
The encoder pipeline aims to compress the primary 2D-reconstruction $\textbf{Y}' \in \mathbb{R}^{B \times 1 \times A \times L}$ in such a way as to retain the structural information from the reconstruction while discarding the noise. The first three stages of the ResNet-34 network are used as the backbone of this feature space encoder but with SE-Attentions embedded within each stage. The primary reconstruction $\textbf{Y}'$ is passed through the first stage of the encoder block, producing $\mathrm{J}_0 $ as
\begin{eqnarray}
    \mathrm{J}_0 = \mathrm{AvgPool2D}\left(\mathcal{E}_0(\textbf{Y}',W_{e0})\right),\;\;\mathrm{J}_0 \in \mathbb{R}^{B \times C \times \frac{A}{2} \times \frac{L}{2}}
\end{eqnarray}
The 2D-ResNet layer is depicted by the notation $\mathcal{E}_0(\cdot)$. Before passing to the next stage, the output feature space $\mathrm{J}_0$ is given attention with the help of an SE block as
\begin{equation}
    \mathrm{J}^{SEatt}_0 = \mathcal{F}_{SEatt}(\mathrm{J}_0, W^{0}_{SEAtt}), \quad \mathrm{J}^{att}_0 \in \mathbb{R}^{B \times C \times \frac{L}{2} \times \frac{A}{2}}
\end{equation}
This process is repeated for the next two stages of the encoder pipeline to compress the feature space as well as lower their spatial dimension. In general, with $\mathrm{J}^{SEatt}_i \in \mathbb{R}^{B \times 2^iC \times 2^{-i-1}A \times 2^{-i-1}L}$ at each stage's output, we can depict each stage as follows:
\begin{equation}
    \mathrm{J}_i = avg_{2\times2}\left(\mathcal{E}_i(\mathrm{J}^{SEatt}_{i-1},W_{ei}) \right), \; \; i \in\{0,1,2\}
\end{equation}
\begin{equation}
    \mathrm{J}^{SEatt}_i = \mathcal{F}_{SEatt}(\mathrm{J}'_i, W^{i}_{SEAtt}), \; \; i \in\{0,1,2\}
\end{equation}
with $\mathrm{J}^{SEatt}_{-1}= \textbf{Y}'$. As a result, three SE-attention weighted encoded features $\mathrm{J}^{SEatt}_0, \mathrm{J}^{SEatt}_1$, and $\mathrm{J}^{SEatt}_2$ are obtained to send to the two decoders.
\vspace{3mm}

\textbf{Regional Decoder Blocks:} 

The foreground and background regions have different modulus values and sizes. Therefore, the remnant noise in the two reconstructed areas are different. Utilizing two dedicated decoders can address the denoising problem more effectively. The final stage encoded feature, given by $ \mathrm{J}^{SEatt}_2 \in \mathbb{R}^{B 
 \times 8C \times \frac{L}{8} \times \frac{A}{8}} $, is taken as input as the first stage of each decoder block. The structure of the decoders is mathematically described below:
\begin{equation}
    \mathrm{D}_i = up_{2 \times 2}\left(\left(\mathcal{F}^{D}_i(D_{i-1},W^{d}_{i}\right)\right),\; i\in\{0,1,2\}    
\end{equation}
\begin{equation}
   \mathrm{D}'_i = concat \left(\mathrm{D}_i, \mathrm{J}_{2-i}\right), \; i\in\{0,1\}
\end{equation}
\begin{equation}
  \mathrm{D}^{SEatt}_i = \mathcal{F}^{i}_{SEatt} \left(\mathrm{D'}_i, W^{i}_{SEatt}\right),  \; i\in\{0,1\}
\end{equation}
with $D_{-1}=\mathrm{J}^{SEatt}_2$. The obtained $D_2 \in \mathbb{R}^{B \times C \times L \times A}$ is trained to be the clean feature in each respective decoder ($D^{FG}_2$: foreground, $D^{BG}_2$: background). To estimate each corresponding modulus, we take $\mathrm{D}_2$ and pass it through a single Conv2D and ReLu processing. The result of this operation is the final YM reconstruction, $Y^{area}$, given by
\begin{equation}
    Y^{area} = \mathcal{F}_{conv,Relu}(\mathrm{D}_2, W_{conv}),\;\; Y^{area} \in \mathbb{R}^{B \times 1 \times A \times L}
\end{equation}
Since we have two decoders, we can obtain the final reconstructions for the FG (foreground) and BG (background) regions using $\mathrm{D}^{FG}_2$ and $\mathrm{D}^{BG}_2$, respectively, as shown below:
\begin{equation}
Y^{FG} = \mathcal{F}_{conv,Relu}^{FG}(\mathrm{D}^{FG}_2, W^{FG}_{conv}),\; Y^{FG} \in \mathbb{R}^{B \times 1 \times A \times L}
\end{equation}
\begin{equation}
Y^{BG} = \mathcal{F}_{conv,Relu}^{BG}(\mathrm{D}^{BG}_2, W^{BG}_{conv}),\; Y^{BG} \in \mathbb{R}^{B \times 1 \times A \times L}
\end{equation}
The resulting reconstructions $Y^{FG}$ and $Y^{BG}$ only possess the cleaned inclusion and background regions, respectively.

\textbf{Fusion Block:}

Proper registration of the background and foreground estimations can produce the final denoised output $\textbf{Y} \in \mathbb{R}^{B \times 1 \times A \times L}$. However, a straightforward addition is not ideal, as the predicted $Y^{FG}$ and $Y^{BG}$ may have some small degree of spatial overlap in their respective boundaries. To achieve a smooth transition between the region boundaries, the features $\mathrm{D}^{FG}_2\in \mathbb{R}^{B \times C \times A \times L}$ and $\mathrm{D}^{BG}_2\in \mathbb{R}^{B \times C \times A \times L}$ are used to effectively fuse the clean areas.

$\mathrm{D}^{FG}_2$ and $\mathrm{D}^{BG}_2$ are concatenated and the channels are gradually reduced using convolution layers, as can be seen from figure \ref{post_denosing_block}. A direct reduction of the channels from $(C+C)$ to $1$ might impede the selectivity of the regional features, hence the gradual channel reduction. The process is described as
\begin{equation}
D^{FG,BG} = BN\left(concat\left(\mathrm{D}^{FG}_2, \mathrm{D}^{BG}_2\right), W^{Fu}_{0}\right)
\end{equation}
\begin{equation}
Y_{pre}^i = \mathcal{F}_{conv,Relu}^i(Y_{pre}^{i-1} , W^{Fu}_{i}), \; i \in \{0,1,2\}
\end{equation}
with $Y_{pre}^{-1}=D^{FG,BG}$. At the ending stage of the above operations, the denoised modulus mapping, \textbf{Y}, is obtained, i.e., $Y_{pre}^{2}=\textbf{Y}$. 

For generating the mask $\textbf{M} \in \mathbb{R}^{B \times 1 \times A \times L}$, the feature stack prior to the denoised output is taken, which is $Y_{pre}^{1}\in \mathbb{R}^{B \times C \times A \times L}$. After proper supervision from $Y^{FG}$ and $Y^{BG}$, $Y_{pre}^{1}$ will possess distinguishing spatial information to isolate the inclusion area. To extract this regional information and map it to a segmentation mask, $\textbf{M}$, the following step is performed:
\begin{equation}
\textbf{M} = \mathcal{F}_{conv,Sigmoid}^4(\mathcal{F}_{conv,Relu}^3(Y_{pre}^1 , W^{Fu}_{3}), W^{Fu}_{4})
\end{equation}

Finally, through our entire proposed pipeline, we obtain: (i) primary reconstruction, $\textbf{Y}'$, (ii) denoised modulus map, $\textbf{Y}$, and (iii) segmentation mask, $\textbf{M}$, which all serve as our main outputs as well as means of supervision. We also obtain two auxiliary features: (iv) foreground or inclusion map, $Y^{FG}$, and (v) background map, $Y^{BG}$ which are used for regional supervision.  

\subsection{Loss Function} 

\begin{figure*}[t]
\centering
\includegraphics[width=.98\textwidth]{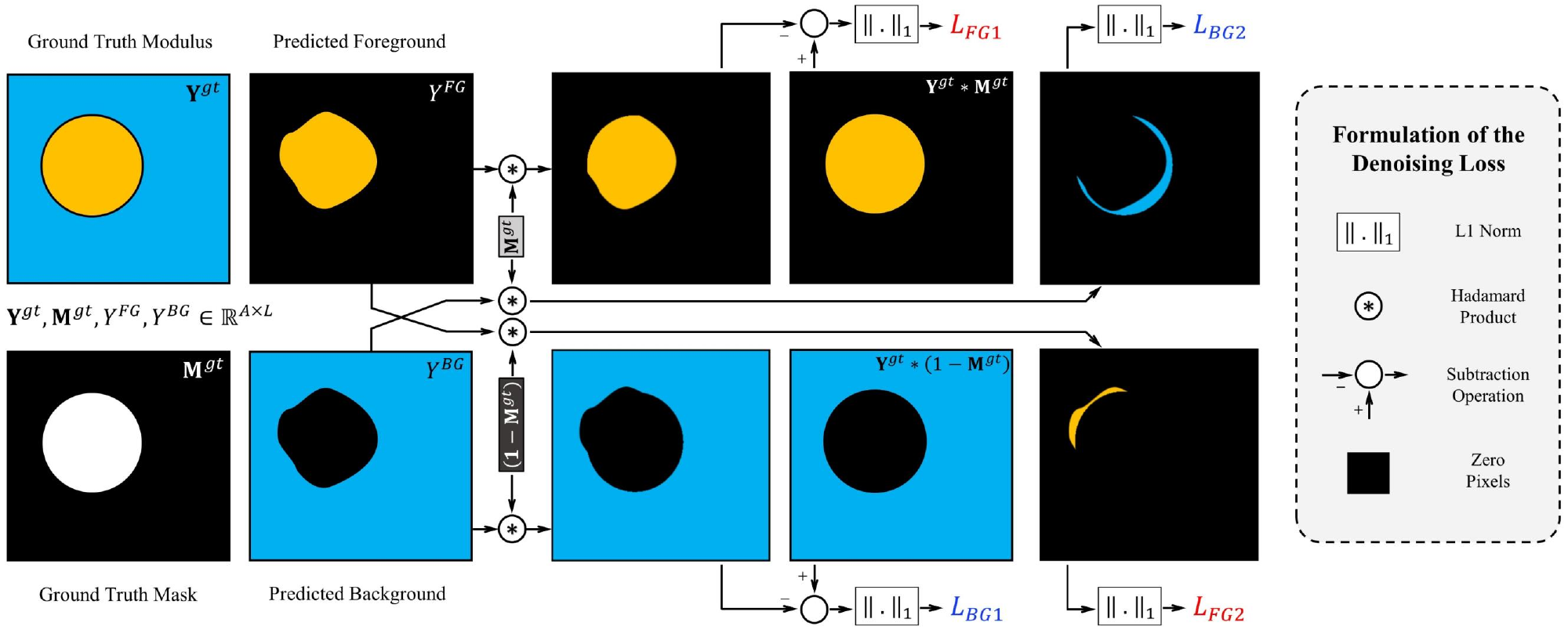}
\caption{Formulation of the denoising loss function, consisting of four components: $L_{FG1}$, $L_{FG2}$, $L_{BG1}$, and $L_{BG1}$.}
\label{denoising_loss_form} 
\end{figure*}

The loss function needs to be specially designed for obtaining expected results from different tasks as well as generalized supervision in limited data scenarios. In the past, reconstruction, classification, and segmentation have been simultaneously done in DL networks for medical tasks, i.e., COVID-19 detection from CT-scan \cite{amyar2020multi}, skin-lesion identification \cite{song2020end}, breast tumors detection in ultrasound \cite{zhou2021multi}, ultrasound SWE stiffness estimation and segmentation \cite{ahmed2021dswe}. Accordingly, we formulate our loss functions that meet our criteria for SWE image reconstruction, denoising, and segmentation. 

Our DL pipeline was designed for two tasks: generating a clean elasticity modulus map of the ROI and creating a prediction mask to separate the foreground from the background. These tasks require the optimization of compatible loss functions. As described in section \ref{Proposed Network Architecture}, the reconstruction network generates only the modulus map; on the contrary, the post-denoiser network performs both denoising and segmentation. The attributes related to high-quality outputs (i.e., clean and low variance outputs with distinct transition between FG and BG) require multiple sub-categories of losses to be minimized; hence the overall losses obtain a compound formation.

\subsubsection{Reconstruction Loss}
The reconstruction network is optimized through the Mean Absolute Error (MAE) between the ground truth and the estimated YM modulus mapping. The MAE loss for reconstruction, functionalized by $\textbf{Y}^{gt}$ and $\textbf{Y}'$, is defined as
\begin{equation}
\begin{split}
    \label{MAE}
    \mathcal{J}_{rec}&=\frac{1}{A.L} \sum_{n=0}^{A-1} \sum_{m=0}^{L-1} \left|{\mathrm{y}}_{n, m}^{gt}-\mathrm{y'}_{n, m}\right| \\
    &= \frac{1}{A.L} \left| \left|{\textbf{Y}}^{gt}-\textbf{Y}'\right| \right|_1
\end{split}    
\end{equation}
where ${y}_{n,m}^{gt}$ and  $y'_{n,m}$ denote the ground truth and estimated modulus value, respectively, at $(n,m)$ spatial coordinate. $\textbf{Y}^{gt}$ and $\textbf{Y}'$ indicate the 2D ground truth and primary reconstruction mappings, respectively. And, $||\cdot||_1$ is used to denote L1-norm.




\subsubsection{Denoising Loss}
The denoising loss function is designed with respect to the two regions: foreground (inclusion) and background. Each region has to be individually cleaned. Additionally, if any predicted foreground resides inside the ground truth background, it is to be penalized (and vice-versa). This ensures that the boundaries of the FG and BG are clean. Keeping these factors in mind, the formulation of the denoising loss is depicted in figure \ref{denoising_loss_form}.

According to figure \ref{denoising_loss_form}, the $L_{FG1}$ and $L_{BG1}$ components provide supervision for cleaning each respective region properly. And, the $L_{FG2}$ and $L_{BG2}$ components aim to mitigate each region leaking into the other. As a result, the denoising loss is given as
\begin{equation}\label{Denoise loss eqn}
    L_{DENOISE}(\alpha_1, \alpha_2)= \alpha_1 L_{FG} +\alpha_2 L_{BG}
\end{equation}
with
\begin{equation} 
    L_{FG}= \frac{1}{A.L}(L_{FG1}+L_{FG2})
\end{equation}
\begin{equation}
    L_{BG}= \frac{1}{A.L}(L_{BG1}+L_{BG2})
\end{equation}
where, for $i=1,2$:
\begin{equation} 
    L_{FGi}= \left| \left| \left[ Y^{FG} -(2-i)\mathbf{Y}^{gt} \right]. \left[ i-1 + (-1)^{i-1} \mathbf{M}^{gt}\right] \right| \right|_1
\end{equation}
and
\begin{equation} 
    L_{BGi}= \left| \left| \left[ Y^{BG} -(2-i)\mathbf{Y}^{gt} \right]. \left[ 2-i + (-1)^{i} \mathbf{M}^{gt}\right] \right| \right|_1
\end{equation}
Here, $\alpha_1$ and $\alpha_2$ are the coupling coefficients for the FG and BG component terms, respectively.



\subsubsection{Fusion Loss}
In order to optimize the fusion block, the final output $\textbf{Y}$ is provided such supervision which will not only penalize the reconstruction error but also increase the shape similarities between $\textbf{Y}^{gt}$ and $\textbf{Y}$. To implement this, we define an MAE-loss and a Normalized-Cross-Correlation (NCC) loss to form our fusion loss, as   
\begin{equation}\label{FUSE loss}
    L_{FUSE}(\beta_1,\beta_2)= \frac{\beta_1}{A.L} \left| \left|\textbf{Y}^{gt}-\textbf{Y}\right| \right|_1 + \beta_2.(1-S_{NCC})
\end{equation}
where, $S_{NCC}$ is given by
\begin{equation}
    S_{NCC}= \frac{\sum_{n=0}^{A-1}\sum_{m=0}^{L-1} y^{gt}_{n,m}\cdot y_{n,m}}{\sqrt{\sum_{n=0}^{A-1}\sum_{m=0}^{L-1} (y^{gt}_{n,m})^2 \cdot \sum_{n=0}^{A-1}\sum_{m=0}^{L-1} (y_{n,m})^2} + \varepsilon}
\end{equation}
The constant $\varepsilon$ is a very small number selected to avoid division by zero. The two losses described in equations (\ref{Denoise loss eqn}) and (\ref{FUSE loss}) clean the regions and merge them optimally on average. However, if there exists some impulsive or outlier noise in $\textbf{Y}'$ prevailing within 1-2 pixels, then that will not be easily reduced in the MAE sense. As such, an additional Total-Variation (TV) loss is considered to smoothen the reconstruction and prevent any sudden noisy jumps in both the axial ($a$) and lateral ($l$) directions, as
\begin{equation}
    L_{TV}= L_{TV}^a+L_{TV}^l
\end{equation}
where
\begin{equation}
    L_{TV}^a=\frac{1}{A-1} \sum_{n=0}^{A-2} \sum_{m=0}^{L-1} \left({\mathrm{y}}_{n, m}-\mathrm{y}_{n+1, m}\right)^2
\end{equation}
and
\begin{equation}
    L_{TV}^l=\frac{1}{L-1} \sum_{n=0}^{A-1} \sum_{m=0}^{L-2} \left({\mathrm{y}}_{n, m}-\mathrm{y}_{n, m+1}\right)^2
\end{equation}
Here, $L_{TV}^a$, $L_{TV}^l$, and $L_{TV}$ denote the axial, lateral, and combined TV loss components, respectively.

\subsubsection{Segmentation Loss}
To optimize the network in producing a segmentation mask, we used the IoU (intersection-over-union) loss function. This supervises the denoiser in detecting the inclusion (FG). If $\mathbf{M}^{gt}$ and $\mathbf{M}$ are defined as the ground truth and predicted binary masks, respectively, the IoU loss function is formed as 

\begin{equation} \label{jaccard}
    L_{\mathrm{IoU}}=1.0- \frac{|\mathbf{M}^{gt} \cap \mathbf{M}|}{|\mathbf{M}^{gt} \cup \mathbf{M}|+\varepsilon}
\end{equation}

\subsubsection{Multi-objective Compound Loss}
The total loss for providing noise resiliency supervision and improved quality enhancement of the reconstructed SWE image is obtained by combining the aforesaid denoising loss, fusion loss, and segmentation loss with appropriate coupling coefficients. We term this loss as $\mathcal{J}_{denoise}$, parameterized by the coefficients $\alpha_1, \alpha_2, \beta_1,\beta_2, \gamma, \mu$:
\begin{equation} \label{Combined_Multi_objective_Loss_equation}
\begin{split}
    \mathcal{J}_{denoise} = L_{DENOISE}(\alpha_1, \alpha_2)&+  L_{FUSE}(\beta_1,\beta_2)\\
    & + \gamma L_{TV} + \mu L_{IoU}
\end{split}
\end{equation}
In this work, the choice of the above coefficients is given below:
\begin{itemize}
    \item $\alpha_1 : \alpha_2$ = Mean ratio between BG and FG pixels in the data
    \item $\beta_1$ = $\kappa.(\alpha_1+\alpha_2)$; $0<\kappa \leq 1$
    \item $\beta_2$, $\gamma$, $\mu$ (iterative) 
\end{itemize}

\section{Experimental Setup}

\subsection{Simulation Data of SWE}
We used COMSOL Multiphysics to design an environment for SW motion propagation in an analogous tissue medium. The "Structural Mechanics" module was previously suggested for this kind of simulation by Ahmed et. al \cite{ahmed2021dswe}. The ARF was simulated with a Gaussian axial force distribution to induce a propagating SW in a 3D Finite Element Model (FEM) study, given by
\begin{equation}\label{ARF}
{ARF}=A_0 \exp \left[-\left(\frac{\left(z-z_0\right)^2}{2 \sigma_z^2}+\frac{\left(x-x_0\right)^2}{2 \sigma_x^2}\right)\right]
\end{equation}
Here, ($z_0$, $x_0$) is the focus point of ARF in the axial and lateral plane; $\sigma_z$ and $\sigma_x$ represent the beam spread in the axial and lateral directions, respectively. According to Palmeri et al. \cite{palmeri2016guidelines}, the push-locale force $A_0$ should be $1000$~$Nm^{-3}$ and the push-duration of $400$~$\mu s$ is selected. This was done to not exceed the maximum tissue displacement of $20$~$\mu m$, due to keeping safety considerations (mechanical index and thermal index) in check.  

Our approach in creating an elasticity map from displacement data is distinct because, we do not follow the phantom generating scheme using a single push outside the ROI or employing multiple pushes like the CUSE technique \cite{song2012comb, song2014fast} or the Supersonic Shear Imaging (SSI) method \cite{bercoff2004supersonic}. Instead, we first divide the ROI into $R$ overlapping regions, as mentioned in equation (\ref{Region_in_vector}). We collect data from each region individually with separate ARFs. Therefore, a total of $R$ pushes are done separately to estimate the stiffness of a single ROI. Each ARF is provided at a lateral offset of $4$~$mm$ from the desired region as blind zones prevail near the push locale. 

\begin{table}[h]
\centering
\caption{Simulation parameters for shear wave generation}
\begin{tabular}{cc}
\hline
\textbf{Parameters} & \textbf{Value} \\
\hline
ARF intensity, $A_0$ & $2 \times 10^5 \ \mathrm{N} / \mathrm{m}^3$ \\
\hline
$\sigma_x, \sigma_y$ & $0.44 \ \mathrm{mm}, 8.00 \ \mathrm{mm}$ \\
\hline
$x_0, z_0$ & Variable \\
\hline
\multirow{2}{*}{Medium} & Nearly incompressible \\
 & linear, isotropic, elastic solid \\
\hline
Poisson's ratio, $v$ & 0.499 \\
\hline
Density, $\rho$ & $1000 \ \mathrm{kg} / \mathrm{m}^3$ \\
\hline
ARF excitation time & $400 \ \mu \mathrm{s}$ \\
\hline
Wave propagation time & $8 \ \mathrm{ms}$ \\
\hline
FEM size & $38\ \mathrm{mm} \times 40 \ \mathrm{mm}$ \\
\hline
Mesh element & Triangular \\
\hline
\end{tabular}

\end{table}
We applied a low-reflecting boundary condition to the edges of our simulated phantoms' Field of View (FOV) to minimize reflection artifacts, but we did not impose this condition on the inclusion boundaries since it is impossible to completely eliminate reflections in real-world situations.

We generated a total of 1380 bi-level phantom datasets, which we then divided into three sets for training, validation, and testing to evaluate the efficacy of our network. The data generation was performed using the COMSOL-MATLAB interface. We varied important parameters, i.e., inclusion diameter, position, stiffness, and background stiffness randomly. Finally, we provided $R=4$ ARFs separately and collected the motion data only from the $R=4$ overlapping regions. In the simulation environment, this process was carried out as follows:

\begin{itemize}
\item An ROI of dimensions ($17.5$~$mm \times 25.7$~$mm$) was selected, and its position was fixed in relation to the FOV of dimensions ($38$~$mm \times 40$~$mm$).
\item An inclusion with a random diameter ranging from $3$~$mm$ to $12$~$mm$ was generated, and its position within the ROI was also randomized.
\item The inclusion and background stiffness (kPa) were randomly varied to create a diverse range of phantoms. Notably, the stiffness values for the inclusion ranged from $8$~$kPa$ to $100$~$kPa$, while for the background, they ranged from $10$~$kPa$ to $35$~$kPa$.
\item $R=4$ number of imaging sequences were obtained from $R=4$ ARF pushes for each ROI individually.  The propagating shear wave was tracked in a $25.7$~$mm \times 7$~$mm$ region, situated $4$~$mm$ laterally offset of each push beam. 
\item The tracked data had an axial and lateral resolution of $8$~$pixel/mm$ and $0.7$~$pixel/mm$, respectively. The shape of a single region of the collected data was $20.5~mm \times 7$~$mm$.
\item The imaging framerate, or pulse repetition frequency for tracking the shear wave, was set at $8$~kHz.

\end{itemize}



\subsection{CIRS Phantom Dataset}
A private dataset consisting of 72 cases of SWE imaging data was obtained from Fuji Healthcare, USA. This data was gathered using CIRS049 phantoms and labeled as A, B, C, and D. Each data featured various types of inclusion stiffness (Type 1-4). For each phantom, 18 ROI were available with various inclusion positions. Table \ref{CIRS049} provides a concise overview of the collected CIRS049 phantom data.

\begin{table}[!h]
    \centering
    \caption{Description of CIRS phantom data type}
    \begin{tabular}{c|ccccc}
    \hline
        CIRS049 & \multicolumn{5}{c}{Type (kPa)} \\ \cline{2-6}
         & BG & 1 & 2 & 3 & 4 \\ \hline
        A & 24 & 7 & 12 & 39 & 66 \\ 
        B & 21 & 6 & 9 & 36 & 76 \\ 
        C & 18 & 6 & 9 & 36 & 72 \\ 
        D & 20 & 6 & 9 & 36 & 72 \\ \hline
    \end{tabular}
    \label{CIRS049}
\end{table}


Each ROI underwent four separate imaging sequences, covering four overlapping regions. The lateral push location remained fixed in relation to the ROI, while adjustments were made to the axial position to match various inclusion depths within the phantom. The raw RF-data obtained during imaging underwent motion conversion and then pre-processing to enhance the quality of the data. This resulted in the final SW motion data samples. 

\begin{figure}[t]
  \centering
  \includegraphics[width=0.48\textwidth]{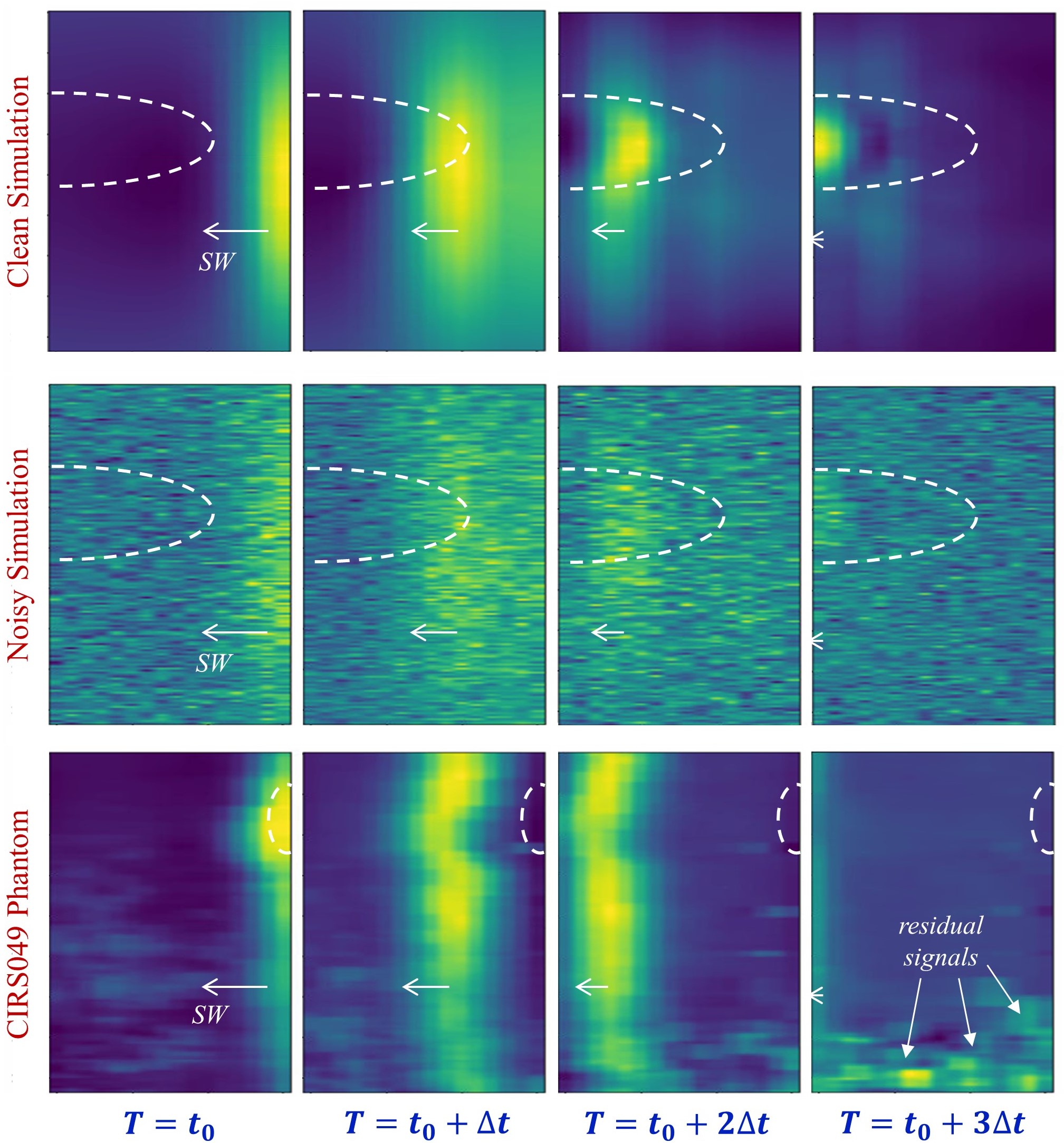}
  \caption{SW showcase of the dataset. The red-dotted line indicates an inclusion. Each SW depiction of a dataset is $\Delta t$ time-frame spaced from each other.}
  \label{data_SW} 
\end{figure}

To illustrate both the simulation as well as CIRS049 phantom dataset, some temporal frames have been shown in figure \ref{data_SW}. It can be seen that the clean simulation showcases a very high contrast SW. When Gaussian noise is added to it, the contrast and signal-to-noise (SNR) ratio is lowered significantly. Finally, the CIRS049 data presents a discernible SW transition from the inclusion area; but it also has residual wave quantities in the background long after the SW had passed. To achieve noise resiliency, we aim to overcome the challenges associated with both low SNR and residual values.

\subsection{Training Procedure}

Each simulation and CIRS case contained $R=4$ regions. They have dimensions of $(168 \times 10)$ pixels in the axial and lateral directions which were physically $(20.5$~$mm \times 10$~$mm)$. We selected the initial 70 temporal frames from each dataset and stacked them to form our 3D volumetric displacement data. To facilitate our training, we interpolated the lateral dimension to 16 pixels. As a result, the final input data ($I_{D,k}$) shape presented to our reconstruction network was $(T, A, L) = (70, 168, 16)$.

The temporal frame count was chosen to be 70 to ensure that the propagating shear wave remained visible within the entire FOV of the regions. In case of the simulation data, we generated two labels, one for the mask and another for the modulus image. In order to ensure the versatility of our networks across different stiffness levels on the simulation dataset, we partitioned it in a way that ensured the training, testing, and validation sets had non-intersecting inclusion stiffness values. Since the stiffness was generated randomly after this split, the total simulation training data consisted of 1010 cases per phantom number, with 111 cases for validation and 259 cases for testing. 

The CIRS049 phantom data possesses the challenge of realistic noise, artifacts, and signal distortions as opposed to the simulation data. Since there are limited samples in the private dataset, the Patch-based training was performed in this case. As a result, an input patch of $\mathbf{I}_{\mathrm{Dp}} \in \mathbb{R}^{B \times 1 \times 70 \times 63 \times 10}$ was mapped to a 2D reconstruction $\textbf{Y}'_p \in \mathbb{R}^{B \times 1 \times 21 \times 4 }$.  Additionally, only the limited samples will consist of limited noise conditions. Therefore, we incorporate the simulation data with the CIRS049 phantoms to be trained simultaneously. We set the training regime in such a way that each epoch of training will be faced with CIRS049 phantom data samples. This is done so that the models do not get over-fitted to the larger simulation data. The patches $\textbf{Y}'_p$ are overlaid on top of each other using a spatial Tukey window to obtain a complete reconstruction of $\textbf{Y}'$.


All our experiments were performed utilizing the PyTorch Deep Learning Framework. We performed normalization on the modulus image, $\textbf{Y}'$  and $\textbf{Y}$, by using the $100$~$kPa$ value as a reference. The input volumetric data, $\textbf{I}_{D,k}$ or, $\textbf{I}_{Dp,k}$ $(k \in \{0,1,...,R-1\})$, underwent min-max normalization. The training of the network was executed on a GeForce RTX 4070 GPU. The primary reconstruction network and the post-denoiser were set to be trained at batch sizes of 8 and 16, respectively. We initially set the learning rate to $10^{-3}$ and used ADAM as our optimizer. We used a reduced learning rate on a plateau scheduler to manage the learning rate dynamically. This scheduler decreased the learning rate by 20\% if the validation loss did not improve for 5 consecutive epochs. The choices for the coefficients from the multi-objective compound loss were: $\kappa=0.5, \beta_2 = 50, \gamma=10, \mu= 1.0$. We conducted all experiments for 150 epochs to ensure convergence on the training and validation sets.

\subsection{Evaluation Metrics}
We performed the quantitative evaluation of our proposed method using the following quality metrics: (i) Peak-Signal-to-Noise-Ratio (PSNR), (ii) Contrast-to-Noise-Ratio (CNR), (iii) Structural Similarity Index (SSIM), (iv) Mean Absolute Error (MAE), (v) Intersection over Union (IoU), (vi) F1-Score, (vii) Hausdorff Distance (HD), (viii) Average Symmetric Surface Distance (ASSD).

\textbf{(i) PSNR:} The PSNR index is commonly used for measuring the reconstruction quality of an image. It is an indicator of the degree of error in the reconstruction. It is defined as
\begin{equation}
    \label{psnr}
    \mathrm{PSNR} = - 10 \cdot \log _{10}\left(\mathrm{MSE}\right)
\end{equation}
\begin{equation}
    \label{MSE}
    \mathrm{MSE}\;\left(\mathcal{I}^N, \hat{\mathcal{I}}^N\right)=\frac{1}{A.L} \sum_{n=0}^{A-1} \sum_{m=0}^{L-1} \left[\mathcal{I}_{n,m}^N-\hat{\mathcal{I}}_{n,m}^N \right]^2
\end{equation}
where, 
\begin{equation} \label{PSNR_normalize}
    \mathcal{I}^N=\frac{\mathcal{I}}{max\left(\mathcal{I}\right)}\;\;\mbox{and}\;\; \hat{\mathcal{I}}^N=\frac{\hat{\mathcal{I}}}{max\left(\hat{\mathcal{I}}\right)}  
\end{equation}
Here, $\mathcal{I}$ and $\hat{\mathcal{I}}$ are the ground-truth 2D modulus image and the estimated 2D modulus image, respectively.

\textbf{(ii) CNR:} The CNR index calculates the contrast between foreground and background with respect to the background standard deviation. It is calculated as
\begin{equation}
    \label{CNR}
    \mathrm{CNR}=20 \log _{10}\left(\frac{|\mu_{FG}-\mu_{BG}|}{\sigma_{BG}}\right)
\end{equation}
where $\mu_{FG}$, $\mu_{BG}$, and $\sigma_{BG}$ are the foreground (inclusion) mean stiffness, the background mean stiffness, and the background standard deviation, respectively.

\textbf{(iii) SSIM:} The SSIM index quantitatively analyzes the perceptual quality between two images. It is defined as
\begin{equation}
    \label{SSIM}
  \mathrm{SSIM}\left(\mathcal{I}, \hat{\mathcal{I}}\right) = \frac{{\left(2\mu_\mathcal{I}\mu_{\hat{\mathcal{\mathcal{I}}}} + \varepsilon_1\right)\left(2\sigma_{\mathrm{cov}} + \varepsilon_2\right)}}{{\left(\mu_\mathcal{I}^2 + \mu_{\hat{\mathcal{I}}}^2 + \varepsilon_1\right)\left(\sigma_{\mathcal{I}}^2 + \sigma_{\hat{\mathcal{I}}}^2 + \varepsilon_2\right)}} 
\end{equation}
where $\mu_\mathcal{I}$ and  $\mu_{\hat{\mathcal{I}}}$ represent the mean of the original and reconstructed images, respectively, $\sigma_{\mathcal{I}}$ and  $\sigma_{\hat{\mathcal{I}}}^2$ represent the standard deviation of the original and reconstructed images, respectively, $\sigma_{\mathrm{cov}}$ denotes the covariance between them. The small constants $\varepsilon_1$ and $\varepsilon_2$ are used to avoid division by zero.

\textbf{(iv) MAE:} The MAE index is also used to assess the reconstruction quality of the modulus image. A higher MAE value indicates a less accurate estimation of tissue stiffness compared to the ground truth. The inclusion (FG) and background (BG) regions are inspected individually. The MAE function is defined in equation (\ref{MAE}).

\textbf{(v) IoU:} The IoU metric is one of the noted metrics for observing the segmentation performance. The IoU loss has been mentioned in equation (\ref{jaccard}). By taking (1-$L_{IoU}$), the performance metric is obtained as

\begin{equation}
    \label{IoU_metric}
  \mathrm{IoU}(\textbf{M}^{gt}, \textbf{M}) = \frac{|\textbf{M}^{gt} \cap \textbf{M}|}{|\textbf{M}^{gt} \cup \textbf{M}|+\varepsilon}
\end{equation}


\textbf{(vi) F1-Score:} The F1-score is calculated from the harmonic mean between the precision and recall metrics \cite{buckland1994relationship} from any prediction. We utilize this metric for observing the segmentation performance. It can also be calculated directly using true-positive (TP), false-positive (FP) and false-negative (FN) values as

\begin{equation}
    \label{F1-score}
     \mathrm{F1} = \frac{2\cdot TP}{2\cdot TP+FP+FN}
\end{equation}

\begin{table*}[t]

\caption{Quantitative Comparison among the all test cases from different Datasets [$\uparrow$: higher is better, $\downarrow$: lower is better]}
    \centering
    \label{Test cases result comparison}
    \begin{tabular}{clccccccc}
    \hline
        \textbf{Data} & 
        \textbf{Method} &
        \textbf{MAE}$\downarrow$ & \textbf{MAE}$\downarrow$ & \textbf{CNR}$\uparrow$  & \textbf{PSNR}$\uparrow$ & \textbf{PSNR}$\uparrow$ & \textbf{PSNR}$\uparrow$ & \textbf{SSIM}$\uparrow$ \\

         \textbf{Structure} &   &\textbf{(FG)} & \textbf{(BG)} &  &  & \textbf{(FG)} & \textbf{(BG)} &  \\ 
         &  &\textbf{[kPa]} & \textbf{[kPa]} & \textbf{[dB]}  & \textbf{[dB]} & \textbf{[dB]} & \textbf{[dB]} &  \\ \hline

        \multirow{4}{*}{}\textbf{Simulation} &  DSWE-Net \cite{ahmed2021dswe} & 3.70  &2.15 & 29.30 & 21.28 & 18.04 & 22.49 & 0.932  \\ 
        (SNR: $\infty$~dB) &  Neidhardt et al. \cite{neidhardt2022ultrasound} &0.76 & 0.15  & 41.04 &29.48&27.78 &29.09 & 0.996 \\ 
        &   \textbf{Ours} ($\mathbf{Y}'$)  &0.83 & 0.40 & 39.53  & 27.95 & 20.54 & 28.56 & 0.988 \\ 
        \textbf{} &  \textbf{Ours} ($\mathbf{Y}$)  & 0.95 & 0.19 & 42.58 & 33.61 & 24.82 & 34.18 & 0.997 \\ 
        \hline

        \multirow{4}{*}{}\textbf{Simulation} & DSWE-Net \cite{ahmed2021dswe} &4.67 &2.45 & 25.85 & 19.17 & 16.08 & 21.33 & 0.911   \\ 
        (SNR: 11~dB) &Neidhardt et al. \cite{neidhardt2022ultrasound} & 3.22 & 1.03  & 28.97  &19.96&20.02 & 19.32& 0.937  \\  
        \textbf{} & \textbf{Ours} ($\mathbf{Y}$) & 1.52 & 0.31 & 43.19 & 32.66 & 23.97 & 33.31 & 0.996 \\ 
        \hline

       \multirow{4}{*}{}\textbf{CIRS049} &DSWE-Net  \cite{ahmed2021dswe} & 8.51& 2.40& 26.06 & 15.96 & 12.92 & 15.91 & 0.899   \\ 
        \textbf{Phantoms} & Neidhardt et al. \cite{neidhardt2022ultrasound} &8.00 &1.31  & 25.29 &16.30&12.08&16.09 &  0.918 \\ 
        
        \textbf{} &\textbf{Ours} ($\mathbf{Y}$) & 4.73 & 1.05 & 36.88 & 22.44 & 16.69 & 19.23 & 0.943 \\ \hline

    \end{tabular}
\end{table*}

\textbf{(vii) Hausdorff Distance (HD) \cite{huttenlocher1993comparing}:} The HD index determines the maximum distance between two sets of surfaces. The differences between two sets of surface pixels are calculated using Euclidean distance. The maximum difference results in the HD index. We want HD to be as low as possible, making it a minimizing metric. If we assume $\mathcal{S}(M^{gt})$ and $\mathcal{S}(M)$ to be the surface pixels of the ground and prediction mask, respectively, and $||\cdot||_e$ to indicate Euclidean distance, the metric can be defined as
\begin{equation}
    \begin{split}
        \mathrm{HD}\left(\textbf{M}^{gt}, \textbf{M}\right) =  \max & \bigg \{   \max_{s_1 \in \mathcal{S}(\textbf{M}^{gt})} d(s_1,\mathcal{S}(\textbf{M})), \\
        & \max_{s_2 \in \mathcal{S}(\textbf{M})} d(s_2,\mathcal{S}(\textbf{M}^{gt})) \bigg \}        
    \end{split}     
\end{equation}
where,
\begin{equation}
     d\left(u,\mathcal{S}(Z)\right) = \min_{s_z \in \mathcal{S}(Z)} \left | \left| u - s_z  \right | \right|_e
\end{equation}

\textbf{(viii) Average Symmetric Surface Distance \cite{heimann2009comparison}:} HD finds the worst-case scenario of distances between the two sets of mask surfaces. We can also use the average symmetric surface distance (ASSD) to determine the overall mask surface gaps in the test cases. Being a minimizing metric, it uses the number of pixels, $n(M^{gt})$ and $n(M)$, of the masks to normalize cross-surface distances. ASSD is calculated as
\begin{equation}
    \begin{split}
        \mathrm{ASSD}\left(\textbf{M}^{gt}, \textbf{M}\right) = \frac{1}{N}& \bigg [   \sum_{s_1 \in \mathcal{S}(\textbf{M}^{gt})} d(s_1,\mathcal{S}(\textbf{M})) \\
        & + \sum_{s_2 \in \mathcal{S}(\textbf{M})} d(s_2,\mathcal{S}(\textbf{M}^{gt})) \bigg ]        
    \end{split}     
\end{equation}
where,
\begin{equation}
    N = n(\textbf{M}^{gt}) + n(\textbf{M})
\end{equation}


\section{Results}

In this section, we conduct a comprehensive comparison of our approach with previously reported deep learning methods, such as Ahmed et al. \cite{ahmed2021dswe} and Neidhardt er al. \cite{neidhardt2022ultrasound},  both qualitatively and quantitatively, using simulation and experimental CIRS phantom data.

\subsection{Performance Evaluation: Simulation Data} \label{simulation_result_section}

Simulation stiffness maps were generated on both clean simulated displacement data and with the addition of Gaussian white noise at 11~dB SNR. Our pipeline was trained following equations (\ref{YM1}), (\ref{YM2}) and (\ref{YM3}) in this case. We examined and evaluated the performance of the deep learning models under a range of challenging conditions by varying the inclusion size, stiffness, and placement in relation to the ARF. The purpose of altering the inclusion size was to see how the network responded when confronted with small and large inclusions. Furthermore, we varied the stiffness of the inclusions so that they were either stiffer or less rigid than the surrounding values. This allowed us to see if the network can not only give stiffness estimates but also determine where the inclusion is, regardless of relative stiffness.  

\begin{figure*}[t]
  \centering
  \includegraphics[width=0.92\textwidth]{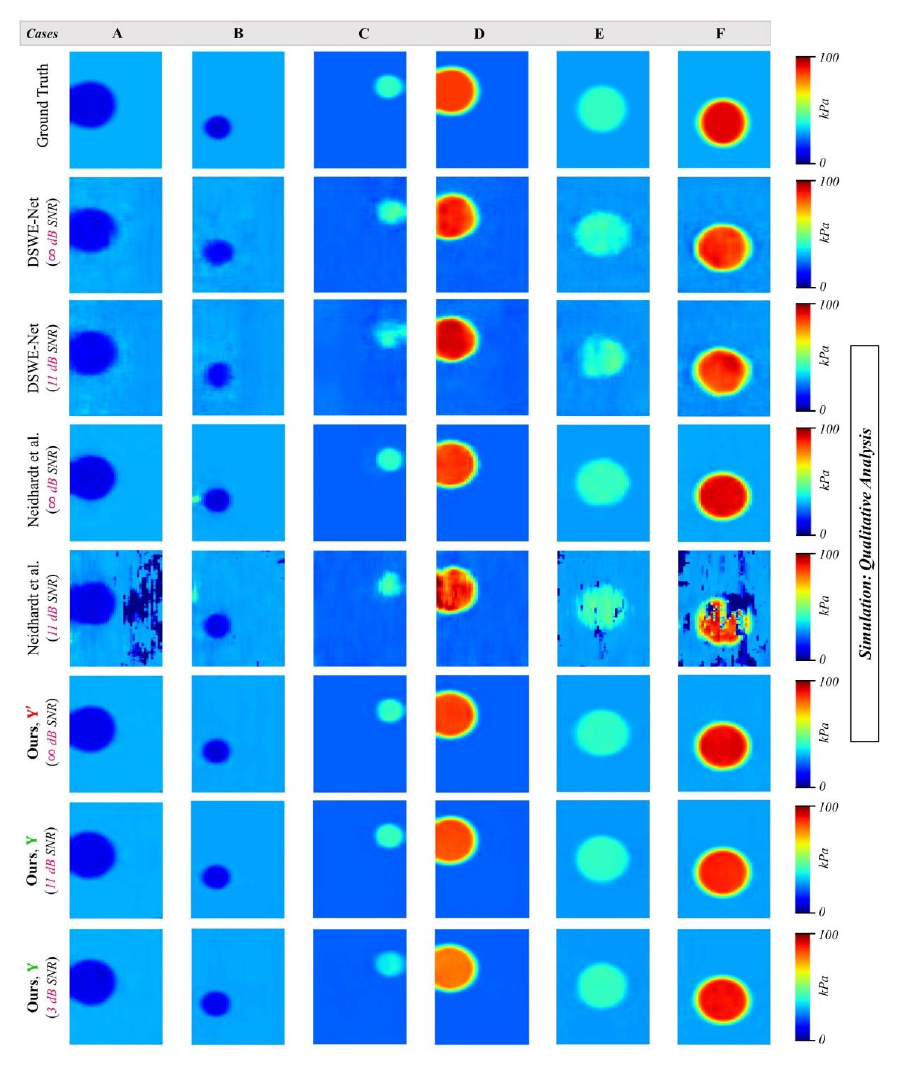}
  \caption{Qualitative reconstruction samples for the simulation test phantoms. The top row shows the ground truths of the six separate phantoms. Results from DSWE-Net \cite{ahmed2021dswe} and Neidhardt et al.\cite{neidhardt2022ultrasound} are produced from 0~dB and 11~dB SNR simulation data. Our method generates results from $\infty$~dB and 3~dB SNR simulation data structure.}
  \label{simu_result_comp} 
\end{figure*}


The quantitative comparison is presented in table \ref{Test cases result comparison}. DSWE-Net \cite{ahmed2021dswe} exhibits significantly higher foreground (FG) and background (BG) MAE at $\infty$ SNR compared to the method proposed by Neidhardt et al. \cite{neidhardt2022ultrasound}. In clean simulation data, Neidhardt et al. achieve very decent FG MAE (0.76 kPa) and BG MAE (0.15~kPa), beating our primary reconstruction $\mathbf{Y}'$ (FG 0.83~kPa, BG 0.40~kPa). Our denoised outputs $\mathbf{Y}$ have slightly increased the FG MAE but cleaner background (FG 0.95~kPa, BG 0.19~kPa). This is a small trade-off in the case of using the post-denoiser in noiseless data. Our $\mathbf{Y}'$ output in clean data demonstrates the high CNR and PSNR at 39.53~dB and 27.95~dB, respectively, under Neidhardt et al. \cite{neidhardt2022ultrasound} at 41.04~dB and 29.48~dB, respectively. This indicates the overall high smoothness of our reconstruction as well as that from Neidhardt et al. \cite{neidhardt2022ultrasound}. Furthermore, the FG PSNR of our denoised output $\mathbf{Y}$ exhibits better performance than $\mathbf{Y}'$ (24.92~dB>20.54~dB), although having worse FG MAE. This implies that our primary reconstruction possessed some outlier peak values in the foreground, which the denoiser eliminated. Based on the preference of having cleaner reconstructions at the expense of slightly increasing the FG error in noiseless data, a user may choose between $\mathbf{Y}'$ and $\mathbf{Y}$.

\begin{figure*}[t]
  \centering
  \includegraphics[width=0.9\textwidth]{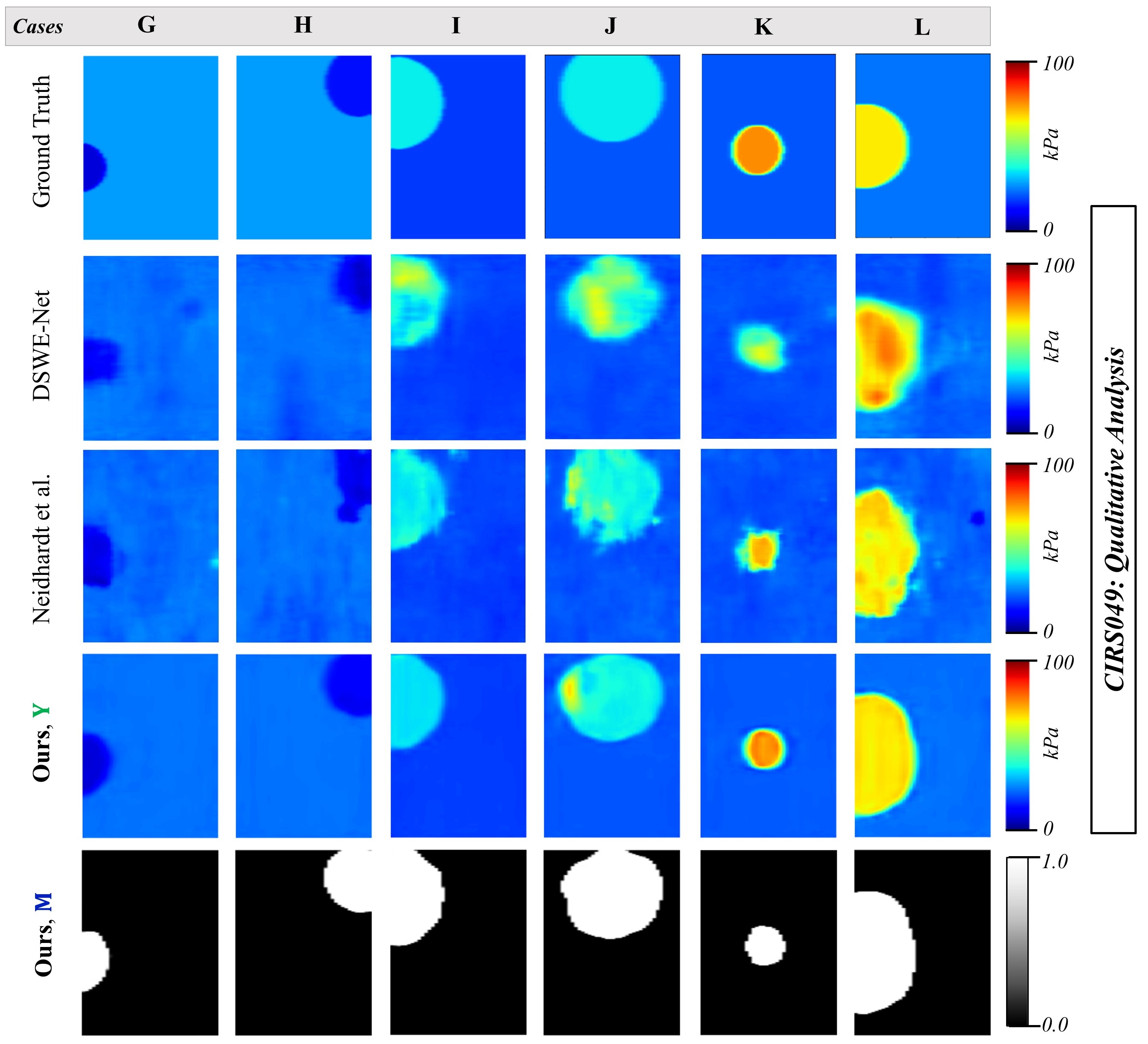}
  \caption{2-D YM image reconstruction for the single inclusion simulation test phantoms. The top row shows the ground truths of the four separate phantoms, the second row the outputs from Neidhardt et al. \cite{neidhardt2022ultrasound} paper, and the third and fourth rows show outputs from DSWE-Net \cite{ahmed2021dswe} and our method respectively.}
  \label{cirs_result_comp} 
\end{figure*}




Analyzing foreground (FG) and background (BG) PSNR separately highlights that Neidhardt et al. follow a similar trend to the FG and BG MAE. They achieve superior FG and BG PSNR at 27.78~dB and 29.09~dB, respectively, exceeding ours (20.54~dB and 28.56~dB). The outlier peaks of our primary reconstructions lowered the FG PSNR due to it using normalization of the image (see equations (\ref{PSNR_normalize})). Furthermore, DSWE-Net \cite{ahmed2021dswe} performs suboptimal among the three methods (FG MAE: 18.04 dB, BG MAE: 22.49 dB). This proves that Neiderhart et al. \cite{neidhardt2022ultrasound} perform on average best in noiseless simulation data. In terms of SSIM, both our method and Neiderhart et al. show similar results (SSIM:  0.997 and 0.996), while DSWE-Net performs poorly (SSIM: 0.932) in comparison.

To assess the robustness of different networks in realistic conditions, we introduced additive Gaussian noise to the simulation data at 11 dB SNR. Neidhardt et al. \cite{neidhardt2022ultrasound} experience a significant increase in MAE (FG: 3.22~kPa, BG: 1.03~kPa), proving the vulnerability of their method to noisy data. DSWE-Net \cite{ahmed2021dswe} degrades a lesser amount (FG MAE: 4.67~kPa, BG MAE: 2.45~kPA), but still falls short. In contrast, our pipeline is explicitly designed for noise resilience, yielding lower mean MAE (FG MAE: 1.52~kPa, BG MAE: 0.31~kPa) than the other methods. This demonstrates the ability to produce accurate and high-quality stiffness maps from noisy data of our pipeline, outperforming previous methods that struggle at the same noise level. 



Our pipeline maintains high PSNR and CNR values in noisy data due to the denoiser's ability to smooth output and reduce standard deviation. The CNR and PSNR outcomes are over 14~dB and 12~dB, respectively, better than DWSE-Net \cite{ahmed2021dswe} and Neidhardt et al. \cite{neidhardt2022ultrasound}. DSWE-Net \cite{ahmed2021dswe} and our technique (denoiser) both pass the entire ROI as inputs and do not fall too short from clean to noisy data. However, Neidhardt et al. \cite{neidhardt2022ultrasound} experience a drastic performance drop in noisy conditions. This proves that being a point-wise DL technique, it declines in handling noise (performance degradation, PSNR: -9.52~dB, PSNR (FG): -7.76~dB, PSNR (BG): -9.77~dB). As for DSWE-Net, although the performance degradation is lower (performance degradation, PSNR: -2.11~dB, PSNR (FG): -1.96~dB, PSNR (BG): -1.16~dB) in 11~dB SNR, it cannot handle a higher degree of noise than that. Our method demonstrates noise robustness up to 3~dB SNR (discussed further in section \ref{discussion}).

Figure \ref{simu_result_comp} provides a qualitative representation of our comparative analysis, depicting the six test samples with different inclusion sizes and positions (color-map indicates in kPa). DSWE-Net \cite{ahmed2021dswe} yields the least satisfactory results with low-quality FG (inclusion) and BG reconstructions in noise-free simulation data (SNR: $\infty$~dB). The results deteriorate when noise is introduced into the data. In contrast, both our model and the approach presented by Neidhardt et al. \cite{neidhardt2022ultrasound} perform very well when presented with noise-free displacement data, generating YM maps with minimal error. When noise is introduced at an 11~dB SNR level, the image quality of the spatio-temporal method by Neidhardt et al. \cite{neidhardt2022ultrasound} declines significantly, evident from the results displayed in the fifth row of figure \ref{simu_result_comp}. On the other hand, our pipeline demonstrates robust performance, maintaining a low level of reconstruction errors. Moreover, at even a lower SNR level of 3~dB, our technique maintains the quality of the reconstruction decently (last row of figure \ref{simu_result_comp}). 


Due to being a point-wise DL method, Neidhardt et al. \cite{neidhardt2022ultrasound} get a big advantage at being trained on significantly more data. This enables it to perform better than our DL approach in noise-less conditions to some degree. However, real-world ultrasound data structures are seldom noise-free. Evaluating the performance metrics and image quality, it is clear that our approach outperforms both Neidhardt et al. \cite{neidhardt2022ultrasound} and the DSWE-Net \cite{ahmed2021dswe} in dealing with noise-filled simulated test phantoms, emulating real-world practical data.  

\subsection{Performance Evaluation: CIRS049 Phantom} \label{cirs_result_section}

Performance of our scheme on the private data is analyzed following the equations (\ref{patch_system_0}), (\ref{patch_system_1}), (\ref{patch_system_2}), (\ref{YM2}) and (\ref{YM3}) in sequence, using the patch-based training regime. Although the BG MAE of the DSWE-Net \cite{ahmed2021dswe} and Neidhardt et al. \cite{neidhardt2022ultrasound} methods are low (respectively 2.40~kPa and 1.31~kPa), their FG MAE are noticeably large (respectively, 8.51~kPa and 8.00~kPa). Whereas, our method yields FG and BG MAE of 4.73~kPa and 1.03~kPa, respectively. Neidhardt et al. \cite{neidhardt2022ultrasound} have the benefit of being trained on the biggest possible data amount as it is a point-wise estimation technique; yet, it possesses the lowest CNR and PSNR due to its inability to map realistic noise. In comparison, our patch-based training deals with far less data. However, it maintains a higher CNR and PSNR (36.88 dB and 22.44 dB, respectively). This suggests that patch-wise training is more appropriate for noise robustness and data scarcity. The individual foreground and background PSNRs follow a similar trend to the PSNR.

Figure \ref{cirs_result_comp} displays the visual results of six test cases, which are consistent with the quantitative performance measurements. DSWE-Net \cite{ahmed2021dswe} produces outcomes that retain the shape of the phantoms to some extent, but the estimates are noisy and the FG-BG boundaries are blurry. The FG-BG boundaries from the method of Neidhardt et al. \cite{neidhardt2022ultrasound} are also very noisy (degrading the phantom shapes), although the estimation color-maps are better than those from the DSWE-Net. On the other hand, our results exhibit estimations that are very clean and high quality. Due to the better structural shape of the inclusions in our estimations, the SSIM metric exceeds that of the other methods noticeably (0.943 > 0.918 > 0.892).

\begin{table}[h]
    \small
    \centering
    \caption{Segmentation performance ($\mathbf{M}$ from our method) on test cases of different datasets [$\uparrow$: higher is better, $\downarrow$: lower is better]}
    \label{table:Masks}
    \begin{tabular}{cccccc}
    \hline
    \textbf{Data} & \begin{tabular}[c]{@{}c@{}}\textbf{SNR}\\ \textbf{(dB)}\end{tabular} &\textbf{IoU}$\uparrow$ & \textbf{F1}$\uparrow$ & \textbf{HD}$\downarrow$ & \textbf{ASSD}$\downarrow$ \\ \hline
        Simulation & $\infty$ & 0.951 & 0.939 & 1.13 & 0.116 \\ 
        Simulation & 3 & 0.909 & 0.917 & 1.58 &  0.227 \\
        CIRS049 & - &0.781 & 0.873 & 6.40 &  0.863\\
        \hline
    \end{tabular}
\end{table}

\subsection{Performance Evaluation: Segmentation Mask} \label{mask_result_section}

Table \ref{table:Masks} indicates the segmentation performance of our proposed method, specifically that of the mask $\textbf{M}$ generated from our post-denoiser model. The clean and noisy simulation dataset yields high IoU (0.951 and 0.909, respectively) and F1-scores (0.939 and 0.917, respectively). As such, the resulting segmentation masks are nearly identical to the ground truths, and thus they are not included in the visual results. We also compute the overall HD and ASSD, which are more sensitive to the mask surface compared to the IoU and F1-score. On average, the maximum pixel-level surface gaps between the infinite and 3~dB SNR test cases are quite low (HD: 1.13 and 1.58, respectively). These low values make the ASSD values quite low as well (0.116 and 0.227, respectively).

\begin{figure*}[t]
  \centering
  \includegraphics[width=1\textwidth]{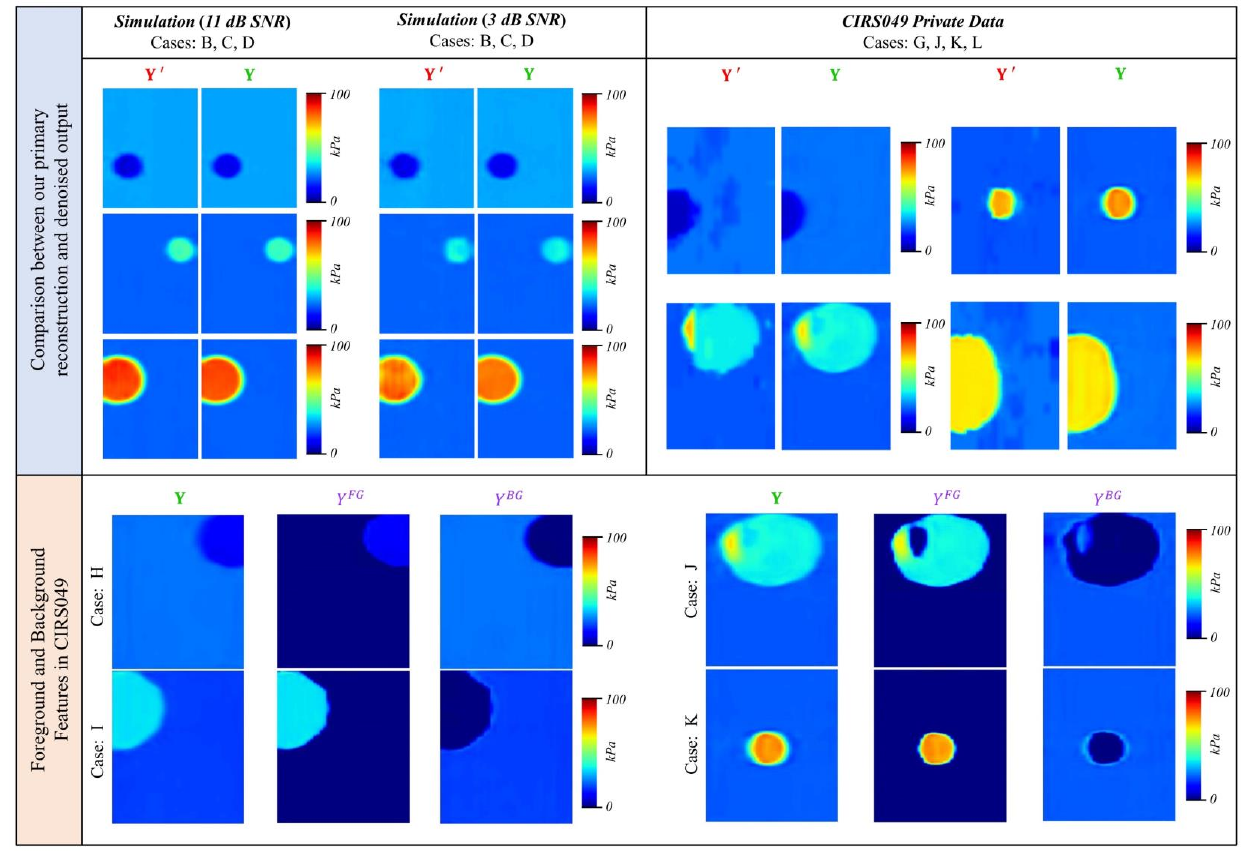}
  \caption{Comparison between the primary reconstruction $\mathbf{Y}'$, foreground feature $Y^{FG}$ and background feature $Y^{BG}$ with respect to the corresponding cleaned output, $\mathbf{Y}$ from our pipeline.  } 
  \label{ALL_intermediate_features} 
\end{figure*}  

For the CIRS049 phantoms, the last row in figure \ref{cirs_result_comp} includes some cases of segmentation masks. The model can identify smaller regions of inclusion in larger areas of background, regardless of their stiffness. The segmentation mask, $\textbf{M}$, is the output from a sigmoid activation function. The mask results shown in figure \ref{cirs_result_comp} are the direct output from that function without being thresholded. This means that the cleaned FG-BG boundaries of the masks possess high confidence. The mean IoU, F1, HD, and ASSD on the overall denoised CIRS049 test samples are respectively 0.781, 0.873, 6.40, and 0.863, as can be seen from table \ref{table:Masks}. The misalignment between the ground truth and estimated masks lowers the IoU to some extent (further explained in section \ref{discussion}). Moreover, the network from Neidhardt et al. \cite{neidhardt2022ultrasound} does not produce segmentation masks or does not present any segmentation strategies to establish a comparison.

\section{Discussion}\label{discussion}

In this paper, a noise-resilient CNN-based deep learning pipeline is proposed with a reconstruction network cascaded with a post-denoiser network. The reconstruction network consists of multi-nested LSTM modules between the encoder and decoder to optimally map the 3D motion data to a 2D modulus estimation. The denoiser contributes to the noise resiliency in the sense that the primary reconstructed images are cleaned to generate high contrast and better-quality images. Sections \ref{simulation_result_section} and \ref{cirs_result_section} provide both qualitative and quantitative evaluations of how our pipeline can produce low-error, low-noise modulus mappings. Section \ref{mask_result_section} provides an analysis of the segmentation mask quality generated from our method. 

\begin{table}[h]

\caption{Performance metric comparison between the primary reconstructions and denoised outputs}
    \centering
    \footnotesize
    \label{Y_prime and Y result comparison}
    \begin{tabular}{ccrrr}
    \hline
        \textbf{Data} & \textbf{Feature} & \textbf{MAE}$\downarrow$ & \textbf{MAE}$\downarrow$ &  \textbf{PSNR}$\uparrow$ \\

         & & \textbf{(FG)[kPa]} & \textbf{(BG)[kPa]} &  \textbf{[dB]}  \\ \hline
        
        \multirow{4}{*}{}\textbf{Simulation} & $\mathbf{Y}'$ & 2.30 & 0.76 & 24.59 \\  
        (3~dB SNR) & $\mathbf{Y}$ & 1.96 & 0.41 &  29.70 \\ 
        \hline

       \multirow{4}{*}{}\textbf{CIRS049} & $\mathbf{Y}'$ & 6.03 & 1.28  & 18.61 \\ 
        \textbf{} & $\mathbf{Y}$  & 4.73 & 1.05  & 22.44 \\ \hline

    \end{tabular}
\end{table}

\begin{table*}[h]
\centering
\caption{Quantitative comparison (mean, median, std. kPa) of various techniques for Simulation test samples depicted in figure \ref{simu_result_comp}}
\small
\label{table:Simu_6_cases}
\begin{tabular}{c|clrrrr}
\hline
\textbf{Type} & \textbf{Cases}   & \multicolumn{1}{c}{\textbf{Method}} & \multicolumn{1}{c}{\begin{tabular}[c]{@{}c@{}}\textbf{FG Mean}\\ \textbf{± STD {[}kPa{]}}\end{tabular}} & \multicolumn{1}{c}{\begin{tabular}[c]{@{}c@{}}\textbf{BG Mean}\\ \textbf{± STD {[}kPa{]}}\end{tabular}} & \multicolumn{1}{c}{\begin{tabular}[c]{@{}c@{}}\textbf{Inclusion}\\ \textbf{Med. {[}kPa{]}}\end{tabular}} & \multicolumn{1}{c}{\begin{tabular}[c]{@{}c@{}}\textbf{Background}\\ \textbf{Med. {[}kPa{]}} \end{tabular}} \\ \hline

\multirow{18}{*}{\rotatebox{90}{Clean (SNR: $\infty$ dB)}}    && DSWENet \cite{ahmed2021dswe}           & 10.75±1.19                                                    & 28.92±0.85                                                    & 10.43                                                          & 28.92                                                           \\  
    && Neidhardt et al. \cite{neidhardt2022ultrasound}  & 9.11±0.69                                                     & 29.91±0.28                                                    & 9.13                                                           & 29.98                                                           \\  
       
&\multirow{-3}{*}{\begin{tabular}[c]{@{}c@{}}A\\ (FG: 9kPa,\\ BG: 30kPa)\end{tabular}}  

& Ours ($\mathbf{Y}$)       & 9.44±0.70                                                     & 29.86±0.26                                                    & 9.25                                                           & 29.77                                                           \\
\cline{2-7}
&& DSWENet \cite{ahmed2021dswe}           & 12.21±1.44                                                    & 28.63±0.74                                                    & 12.02                                                          & 28.63                                                           \\

&& Neidhardt et al. \cite{neidhardt2022ultrasound}  & 8.51±0.41                                                                           & 29.07±1.10                                                                            & 8.56      & 29.01     \\

&\multirow{-3}{*}{\begin{tabular}[c]{@{}c@{}}B\\ (FG: 8kPa,\\ BG: 29kPa)\end{tabular}}  & 

Ours ($\mathbf{Y}$)       & 8.82±0.47                                                     & 28.92±0.35                                                    & 8.83                                                           & 28.92           \\ \cline{2-7}

 && DSWENet \cite{ahmed2021dswe}           & 42.53±2.51                                                    & 22.09±0.52                                                    & 42.61                                                          & 22.15                                                           \\
 
 && Neidhardt et al. \cite{neidhardt2022ultrasound}  & 39.89±0.60                                                                           & 22.27±0.67                          & 39.94  
 & 22.30                  \\

&\multirow{-3}{*}{\begin{tabular}[c]{@{}c@{}}C\\ (FG: 41kPa,\\ BG: 22kPa)\end{tabular}} & Ours ($\mathbf{Y}$)       & 41.71±1.14                                                    & 21.98±0.16                                                    & 41.70                                                          & 21.97                                                           \\ \cline{2-7}
&& DSWENet \cite{ahmed2021dswe}           & 86.29±3.71                                                    & 22.30±0.79                                                    & 87.63                                                          & 22.30                                                           \\

&& Neidhardt et al. \cite{neidhardt2022ultrasound}  & 84.79±1.51                                                    & 22.02±0.23                                                    & 85.02                                                          & 22.00                                                           \\

&\multirow{-3}{*}{\begin{tabular}[c]{@{}c@{}}D\\ (FG: 85kPa,\\ BG: 22kPa)\end{tabular}} & 

Ours ($\mathbf{Y}$)       & 85.48±1.93                                                    & 21.94±0.19                                                    & 85.17                                                          & 21.94                                                           \\
\cline{2-7}

&& DSWENet \cite{ahmed2021dswe}           & 39.90±1.76                                                    & 27.21±0.61                                                    & 39.68                                                          & 27.26                                                           \\

&& Neidhardt et al. \cite{neidhardt2022ultrasound}  & 41.46±0.83                                                                            & 28.03±0.03                                                                            & 41.52                                                                                  & 28.00                                                                                   \\

&\multirow{-3}{*}{\begin{tabular}[c]{@{}c@{}}E\\ (FG: 41kPa,\\ BG: 28kPa)\end{tabular}} & 

Ours ($\mathbf{Y}$)       & 40.93±0.67                                                    & 27.90±0.20                                                    & 40.93                                                          & 27.89                                                           \\
\cline{2-7}

  && DSWENet \cite{ahmed2021dswe}           & 85.60±2.78                                                    & 27.33±0.92                                                    & 85.55                                                          & 27.48                                                           \\
  && Neidhardt et al. \cite{neidhardt2022ultrasound}  & 91.13±0.99 & 29.01±0.29 & 91.28 & 29.00\\
  
&\multirow{-3}{*}{\begin{tabular}[c]{@{}c@{}}F\\ (FG: 91kPa,\\ BG: 29kPa)\end{tabular}} & 

Ours ($\mathbf{Y}$)       & 90.26±1.27                                                    & 28.90±0.25                                                    & 90.53                                                          & 28.90     \\ \hline         

\multirow{18}{*}{{\rotatebox{90}{Noisy (SNR: 11~dB) }}} & \multirow{3}{*}{\begin{tabular}[c]{@{}c@{}}A\\ (FG: 9kPa,\\ BG: 30kPa)\end{tabular}}                                                              & DSWENet \cite{ahmed2021dswe}          & 11.07±1.56            & 28.50±1.00            & 10.64                 & 28.33                 \\  

& & Neidhardt et al. \cite{neidhardt2022ultrasound} & 7.49±4.04 & 28.51±1.65 & 9.02 & 28.93 \\  
                                                   
                        &                                                                       & Ours ($\mathbf{Y}$)      & 9.42±0.56             & 29.82±0.29            & 9.36                  & 29.85                 \\ \cline{2-7} 
                        & \multirow{3}{*}{\begin{tabular}[c]{@{}c@{}}B\\ (FG: 8kPa,\\ BG: 29kPa)\end{tabular}}                                                               & DSWENet \cite{ahmed2021dswe}          & 11.84±1.69            & 28.09±0.89            & 11.24                 & 29.09                 \\  
                        & & Neidhardt et al. \cite{neidhardt2022ultrasound} & 10.10±0.95 & 28.86±1.22 & 9.83 & 28.86 \\
                        &                                                                                       & Ours ($\mathbf{Y}$)      & 10.07±0.49            & 28.90±0.33            & 9.99                  & 28.93                 \\ \cline{2-7} 
                        & \multirow{3}{*}{\begin{tabular}[c]{@{}c@{}}C\\ (FG: 41kPa,\\ BG: 22kPa)\end{tabular}}                                       & DSWENet \cite{ahmed2021dswe}          & 37.37±2.86            & 22.66±0.96            & 37.65                 & 22.59                 \\  
                        & & Neidhardt et al. \cite{neidhardt2022ultrasound} & 42.06±2.29 &22.31±0.63 & 42.53 & 22.23 \\
                        &                                                                                       & Ours ($\mathbf{Y}$)      & 36.23±0.95            & 21.90±0.27            & 36.25                 & 21.89                 \\ \cline{2-7} 
                        & \multirow{3}{*}{\begin{tabular}[c]{@{}c@{}}D\\ (FG: 85kPa,\\ BG: 22kPa)\end{tabular}}                                                                & DSWENet \cite{ahmed2021dswe}          & 90.53±3.00            & 22.40±0.99            & 91.71                 & 22.30                 \\  
                        & & Neidhardt et al. \cite{neidhardt2022ultrasound} & 86.39±5.86  & 22.45±0.63  & 86.53& 22.38 \\
                        &                                                                                       & Ours ($\mathbf{Y}$)      & 82.98±0.98            & 21.90±0.20            & 83.08                 & 21.91                 \\ \cline{2-7} 
                        & \multirow{3}{*}{\begin{tabular}[c]{@{}c@{}}E\\ (FG: 41kPa,\\ BG: 28kPa)\end{tabular}}                                                              & DSWENet \cite{ahmed2021dswe}          & 40.69±3.33            & 26.64±0.84            & 40.44                 & 26.67                 \\  
                        & & Neidhardt et al. \cite{neidhardt2022ultrasound} & 43.10±1.86 & 27.83±2.77 & 43.24 & 28.22 \\
                        &                                                                                       & Ours ($\mathbf{Y}$)      & 41.33±0.49            & 27.69±0.19            & 41.31                 & 27.69                 \\ \cline{2-7} 
                        & \multirow{3}{*}{\begin{tabular}[c]{@{}c@{}}F\\ (FG: 91kPa,\\ BG: 29kPa)\end{tabular}}                     & DSWENet \cite{ahmed2021dswe}          & 85.78±3.45            & 27.04±1.02            & 85.41                 & 27.10                 \\                         & & Neidhardt et al. \cite{neidhardt2022ultrasound} & 81.19±8.84 & 24.51±7.98 & 84.21 & 28.01 \\
                        &                                                                                       & Ours ($\mathbf{Y}$)      & 89.19±0.84            & 28.67±0.34            & 89.36                 & 28.67                 \\ \hline

\end{tabular}
\end{table*}

\begin{table*}[h]
\centering
\caption{Quantitative comparison (mean, median, std. kPa) of various techniques for CIRS049 test samples depicted in figure \ref{cirs_result_comp}}
\small
\label{table:cirs_6_cases}
\begin{tabular}{c|clrrrr}
\hline
\textbf{Type} & \textbf{Cases}   & \multicolumn{1}{c}{\textbf{Method}} & \multicolumn{1}{c}{\begin{tabular}[c]{@{}c@{}}\textbf{FG Mean}\\ \textbf{± STD {[}kPa{]}}\end{tabular}} & \multicolumn{1}{c}{\begin{tabular}[c]{@{}c@{}}\textbf{BG Mean}\\ \textbf{± STD {[}kPa{]}}\end{tabular}} & \multicolumn{1}{c}{\begin{tabular}[c]{@{}c@{}}\textbf{Inclusion}\\ \textbf{Med. {[}kPa{]}}\end{tabular}} & \multicolumn{1}{c}{\begin{tabular}[c]{@{}c@{}}\textbf{Background}\\ \textbf{Med. {[}kPa{]}} \end{tabular}}                   \\ \hline
\multirow{18}{*}{\rotatebox{90}{CIRS049 Phantoms}} & \multirow{3}{*}{\begin{tabular}[c]{@{}c@{}}G\\ (FG: 7kPa,\\ BG: 24kPa)\end{tabular}}  & DSWENet \cite{ahmed2021dswe}          &          12.03±1.32                                                                                         &        23.65±0.88                                                                                          &      12.26                        &             23.75                                 \\
                          &                                                                                       & Neidhardt et al. \cite{neidhardt2022ultrasound} &        9.39±1.16                                                                                          &        23.24±1.15                                                                                       &      9.23                        &             23.26                   \\
                          &                                                                  & Ours ($\mathbf{Y}$)     & 8.01±0.47                                                                                         & 23.70±0.27                                                                                       & 8.00                         & 23.74                       \\ \cline{2-7} 
                          & \multirow{3}{*}{\begin{tabular}[c]{@{}c@{}}H\\ (FG: 12kPa,\\ BG: 24kPa)\end{tabular}} & DSWENet \cite{ahmed2021dswe}          &        8.51±1.49                                                                                          &  23.23±1.16                                                                                                 &            8.18                 &      23.57                                                   \\
                          &                                                                                       & Neidhardt et al. \cite{neidhardt2022ultrasound} &         9.29±1.13                                                                                           &        23.74±1.02                                                                                          &       9.81                        &             23.74                  \\
                          &                                                                    & Ours ($\mathbf{Y}$)     & 11.73±0.96                                                                                        & 23.69±0.25                                                                                       & 12.03                        & 23.74                       \\ \cline{2-7} 
                          & \multirow{3}{*}{\begin{tabular}[c]{@{}c@{}}I\\ (FG: 36kPa,\\ BG: 18kPa)\end{tabular}} & DSWENet \cite{ahmed2021dswe}          &         42.46±8.53                                  &        18.84±0.86                                                                                           &      39.39                        &             18.73                \\
                          &                                                                                       & Neidhardt et al. \cite{neidhardt2022ultrasound} &        35.78±1.60                                                                                          &       18.98±0.98                                                                                          &      35.77                        &             19.00                \\
                          &                                                                 & Ours ($\mathbf{Y}$)      & 34.72±0.56                                                                                        & 18.45±0.26                                                                                       & 34.78                        & 18.48                       \\ \cline{2-7} 
                          & \multirow{3}{*}{\begin{tabular}[c]{@{}c@{}}J\\ (FG: 36kPa,\\ BG: 21kPa)\end{tabular}} & DSWENet \cite{ahmed2021dswe}          &         47.69±7.42                                                                                           &        20.31±0.97                                                                                          &      47.56                        &             20.27                                            \\
                          &                                                                                       & Neidhardt et al. \cite{neidhardt2022ultrasound} &         39.53±4.99                                                                                          &        20.88±0.96                                                                                     &      37.77                        &             20.80                  \\
                          &                                                                & Ours ($\mathbf{Y}$)      & 38.96±6.33                                                                                        & 20.86±0.42                                                                                       & 36.85                        & 20.91                       \\ \cline{2-7} 
                          & \multirow{3}{*}{\begin{tabular}[c]{@{}c@{}}K\\ (FG: 76kPa,\\ BG: 21kPa)\end{tabular}} & DSWENet \cite{ahmed2021dswe}          &         52.62±6.38                                                                                          &        20.17±0.92                                                                                          &     53.39                        &             20.08                                           \\
                          &                                                                                       & Neidhardt et al. \cite{neidhardt2022ultrasound} &      68.77±5.70                                                                                          &                    20.62±1.22                                                                                 &      70.92                        &             20.48                        \\
                          &                                                                 & Ours ($\mathbf{Y}$)      & 75.28±1.20                                                                                        & 21.28±0.33                                                                                       & 75.08                        & 21.30                       \\ \cline{2-7} 
                          & \multirow{3}{*}{\begin{tabular}[c]{@{}c@{}}L\\ (FG: 66kPa,\\ BG: 24kPa)\end{tabular}} & DSWENet \cite{ahmed2021dswe}          &         69.22±5.25                                                                                           &        22.25±2.14                                                                                          &      69.45                        &             21.98                                            \\
                          &                                                                                       & Neidhardt et al. \cite{neidhardt2022ultrasound} &          65.38±2.75                                                                                         &        22.41±2.02                                                                                         &      65.44                        &             22.24                   \\
                          &                                                                   & Ours ($\mathbf{Y}$)      & 66.35±0.95                                                                                        & 23.37±0.39                                                                                       & 66.26                        & 23.37                       \\ \hline
\end{tabular}
\end{table*}



The intermediate feature sets $\{\textbf{Y}', Y^{FG}$, $Y^{BG}\}\in \mathbb{R}^{1 \times A \times L}$ provide insight on how our pipeline is being methodically supervised. The output $\textbf{Y}'$ of the primary reconstruction network is a 2D full ROI image which is fed to the denoiser for refinement using the supervisory multi-objective loss function defined in equation (\ref{Combined_Multi_objective_Loss_equation}). Figure \ref{ALL_intermediate_features} depicts the comparison among the primary reconstructions, $\textbf{Y}'$, and denoised final outputs, $\textbf{Y}$, in both simulation data (SNR: 11~dB, 3~dB) and the private dataset. The $\textbf{Y}'$ features, although accurate in estimations, have a higher standard deviation compared to their $\textbf{Y}$ counterparts. The denoiser network learns the shape information and the modulus distribution of the data. However, it still relies on the estimation from the original reconstruction network, as the denoiser is not introduced to any 3D motion features. This matter is clear in figure \ref{ALL_intermediate_features} (3~dB simulation, third row, case-D foreground). The $\textbf{Y}$ is cleaner than $\textbf{Y}'$, but the color maps are similar ($\textbf{Y}'$: 76.93±3.36~kPa, $\textbf{Y}$: 78.20±0.91~kPa). Whereas, the foreground color map in the case-D ground truth is slightly darker (e.g., 85~kPa). Therefore, any poor $\textbf{Y}'$ reconstruction due to lower than 3~dB SNR data will result in an under- or over-estimated, clean $\textbf{Y}$. Nevertheless, the denoiser network improves upon the primary reconstruction significantly. The overall reconstruction improvement capability of the denoiser is shown in table \ref{Y_prime and Y result comparison}.

The foreground $Y^{FG}$ and background  $Y^{BG}$ were directly supervised using the loss terms $L_{FG}$ and $L_{BG}$, respectively. To investigate whether the corresponding losses have performed in an anticipated manner, we present some test cases from the CIRS049 dataset in figure \ref{ALL_intermediate_features} showing their corresponding $\textbf{Y}$, $Y^{FG}$, and $Y^{BG}$. We see that the $Y^{FG}$ feature contains only the foreground estimations with the background values zeroed out. Similarly, in the $Y^{BG}$ feature, only the background values exist, with foreground values at zero. The Fusion Block takes the denoised features, $D_2^{FG}$ and $D_2^{BG}$, as inputs to perform feature registration and generate a clean output estimation, $\textbf{Y}$. As can be seen from figure \ref{ALL_intermediate_features}, the fusion was performed very decently.

The stiffness estimations of the six individual cases (A–E) shown in figure \ref{simu_result_comp} are presented in table \ref{table:Simu_6_cases}. The cases range from different phantom sizes, FG-BG modulus differences, and inclusion positions. In clean simulation data, the method from Neidhardt et al. \cite{neidhardt2022ultrasound} showcases the best results of mean and median stiffness for each test case. However, our method exhibits comparatively lower BG standard deviations in this case. When a noise level at 11~dB SNR is introduced to the simulation data, our method outperforms that Neidhardt et al. \cite{neidhardt2022ultrasound} in most estimation cases, in terms of mean, median, and standard deviation. This quantitative observation is consistent with the visual results in figure \ref{simu_result_comp}. Our method can perform well up to an SNR level of 3~dB. Whereas, the other two methods were not able to reconstruct from the 3~dB SNR data even after proper training, and therefore results from them were not included. Similarly, the stiffness estimations of the six cases (G-L), illustrated in figure \ref{cirs_result_comp}, are presented in table \ref{table:cirs_6_cases}. In experimental data, our method again outperforms Neidhardt et al. \cite{neidhardt2022ultrasound} for most cases, with very few exceptions. Furthermore, DSWE-net performs the worst across all the datasets. This implies that our DL pipeline is efficient in generating higher-quality data in noise-inflicted synthetic and realistic data compared to the other DL techniques.  




It should be emphasized that the denoiser network aims to best refine the 2D phantom image that arrives from the primary reconstruction network. It does not have information regarding 3D motion propagation or its related features. Therefore, the denoiser cannot mitigate drastic shape changes or huge deviations in reconstructions. If the mean estimate has deviated from the ground truth, then the refined output, although cleaner, will have deviated results as well. To map the $\textbf{Y}'$ as closely as possible to the ground label, the primary reconstruction network must be trained with sufficient data. Also, the denoiser will fix $\textbf{Y}'$ into shapes on which it has been trained (i.e., circular phantoms in this case). Therefore, a denoiser trained in a dataset with irregular phantom shapes will learn to clean differently from a denoiser learning from simulation and CIRS049 phantoms, which will attempt to make the $\textbf{Y}$ outputs more circular or round-shaped (see figure \ref{cirs_result_comp}: row 4).

\begin{figure*}[t]
  \centering
  \includegraphics[width=1\textwidth]{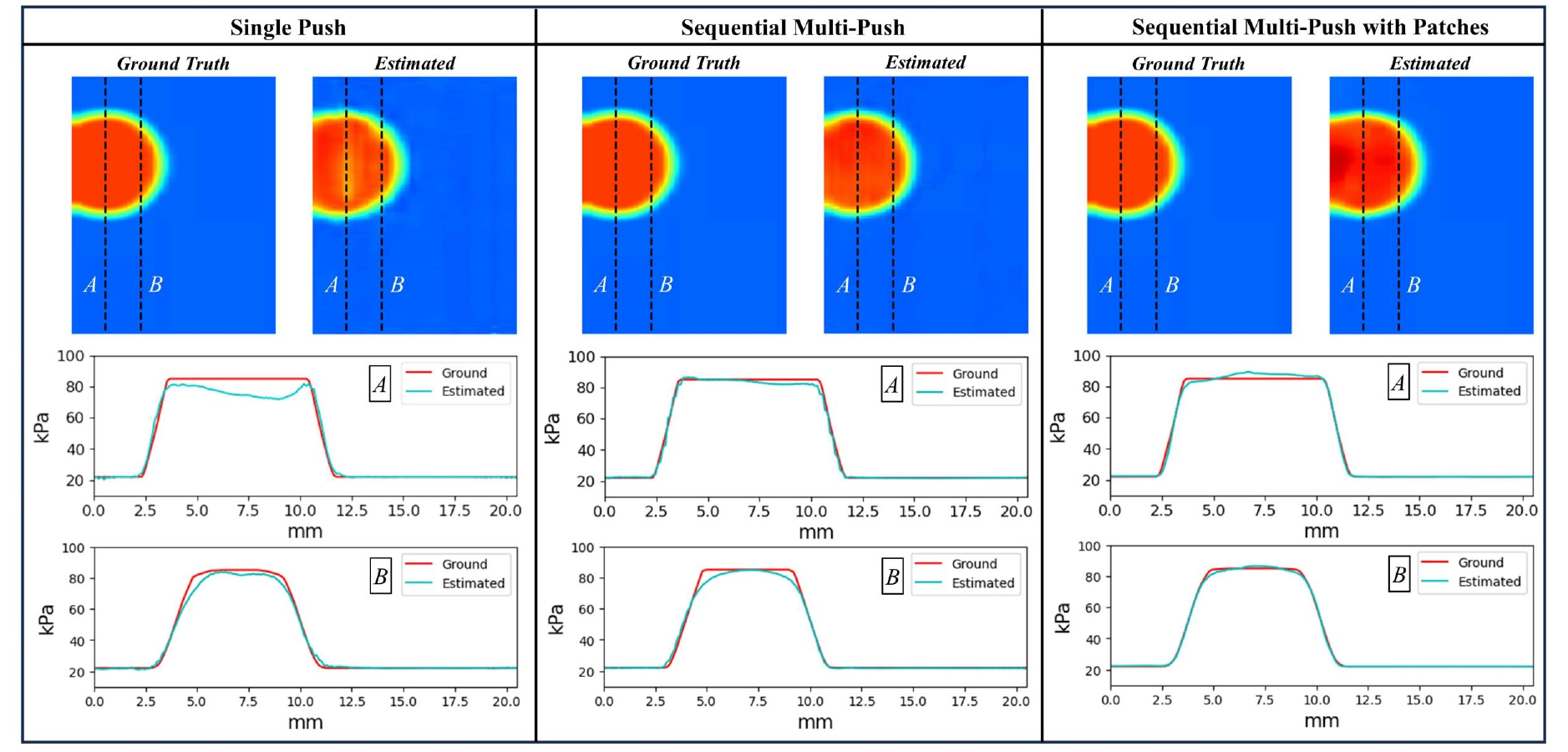}
  \caption{Quality comparisons between training on simulation-based single push, sequential multi-push and sequential multi-push with patched data using axial slices (\textit{A} and \textit{B}).}
  \label{single_multi} 
\end{figure*} 

Another important issue to be addressed is regarding the relatively higher MAE of the inclusions in CIRS049 phantom data (table \ref{Test cases result comparison}: 8.51, 8.00, and 4.73~kPa). This is due to some label issues with the private data. The ground truth modulus maps were provided to be hand-labeled with the CVAT software from B-mode images of the phantoms. The label boundaries were set where a transition was visually seen between the FG and BG regions in the B-mode images. In some B-mode images, the transitions were not visually distinct and the labels were inconsistent with the actual modulus mapping, leading to overlaps in the FG and BG regions. Higher stiffness is found in FG regions, and higher MAE is produced when these overlapping areas exist. We can claim this inconsistency since in a few cases, the models (previously reported and ours) created comparable phantom borders but differed from the ground truth. For instance, in figure \ref{cirs_result_comp}, Case L is reconstructed with a larger inclusion area than the ground truth for all models. Since Case L stiffness values (FG: 66~kPa and BG: 24kPa) are very different, the MAE metric produces a high value when the ground truth BG around the inclusion is overlaid with the estimated FG. Outliers similar to these have increased the overall MAE of the CIRS049 test cases. However, if we look at the mean and median values of Case L within our estimated FG in table \ref{table:cirs_6_cases}, we find that they are, respectively, 66.09~kPa and 66.35~kPa (very near the ground FG). This aspect also explains the relatively low value of the IoU (0.781) for our segmentation.

ROIs for multi-push-based SWE data do not have to be as large as those for single-push data. Because the decayed intensity of one ARF can be compensated by the other ARFs. As a result, the advantage of multi-push data over singular-push data is that ROI counts increase, resulting in increased training samples. Additionally, training in a patched configuration increases training data by an even greater amount. The comparative reconstructions due to these arrangements are demonstrated in figure \ref{single_multi}. The figure illustrates a test sample from simulation (SNR: $\infty$~dB) from our primary reconstructions. Two axial slices (\textit{A} and \textit{B}) between the ground truth and estimated reconstructions have been taken for analysis. Training on single-push data provides noisy outcomes. The average FG and BG MAE from single push data are 1.89~kPa and 0.49~kPa, respectively. The FG MAE from sequential multi-push without and with patches yields close results of 0.83~kPa and 0.88~kPa. However, BG MAE from the patched version (0.14~kPa) outperforms the non-patched configuration (0.40~kPa). Moreover, if we observe the `\textit{B}' slices from figure \ref{single_multi}, demonstrating a thinner part of the inclusion, we see that the patch-based training follows the ground truth value better than the other two. But the slight overestimation in slice `\textit{A}' can result from a lower spatial context, which the non-patched sequential multi-push training possesses better. Therefore, the patch-based training regime should be carried out when ample data is required for appropriate training.




Finally, a crucial point to be mentioned is that the reconstruction effectiveness on noise-added simulation data as well as CIRS049 tissue mimicking experimental data may not fully account for the practical challenges during $in-vivo$ SWE. Clinical tissue medium is seldom as broad region-based homogeneous as experimental ones. Additionally, the noise conditions will differ from the ones we have investigated in this study, affecting the training process. In future, we aim to evaluate as well as tweak our method with clinical SWE data, which is otherwise unavailable to us at present. This will largely improve the generalization capability of our method, which has the potential for real-world deployment.

\section{Conclusion}\label{conclusion}
In this work, we have proposed a Multi-Nested-LSTM embedded CNN network cascaded with a noise-resilient post-denoiser model to generate high-quality SWE reconstructions from simulation and phantom motion data. The Multi-Nested-LSTM embedded CNN acts as the primary reconstruction model to map 3D motion data to 2D modulus images. The subsequent denoiser model is supervised with a multi-objective compound loss consisting of inclusion (foreground) and background mapping terms, a region fusing term, a TV loss term, and an IoU loss term. Such supervision enables the denoiser to take in the primary reconstruction as input and produce not only cleaner images but also corresponding segmentation masks. Our method has been tested on COMSOL-simulated as well as CIRS049 phantoms with different shapes, stiffness, and locations. The resulting estimations are superior to the ones generated from previously reported deep learning SWE estimation techniques: DSWE-Net \cite{ahmed2021dswe} and a spatio-temporal CNN \cite{neidhardt2022ultrasound}. The simultaneously produced segmentation masks are also precise in isolating foreground and background regions, regardless of their stiffness. The performance of our method shows promise and paves the way for investigating reconstruction as well as segmentation capabilities in $in-vivo$ SWE data (i.e., breast, liver, etc.) as a prospective future work.


\bibliographystyle{IEEEtran}



\bibliography{cas-refs.bib}



\end{document}